\documentclass[twocolumn]{aastex61}
\usepackage{natbib}
\usepackage{graphics}

\accepted{ApJ; October 1, 2018}

\shorttitle{Molecular gas in colliding galaxies at $z=1.52$}
\shortauthors{Silverman et al.}

\begin{document}


\title{Concurrent starbursts in molecular gas disks within a pair of colliding galaxies at $\lowercase{z}=1.52$}


\author{J.~D.~Silverman}
\email{john.silverman@ipmu.jp}
\affiliation{Kavli Institute for the Physics and Mathematics of the Universe, The University of Tokyo, Kashiwa, Japan 277-8583 (Kavli IPMU, WPI)}

\author{E.~Daddi}
\affiliation{Laboratoire AIM, CEA/DSM-CNRS-Universite Paris Diderot, Irfu/Service d'Astrophysique, CEA Saclay}

\author{W.~Rujopakarn}
\affiliation{Department of Physics, Faculty of Science, Chulalongkorn University, 254 Phayathai Road, Pathumwan, Bangkok 10330, Thailand}
\affiliation{National Astronomical Research Institute of Thailand (Public Organization), Don Kaeo, Mae Rim, Chiang Mai 50180, Thailand}
\affiliation{Kavli Institute for the Physics and Mathematics of the Universe, Todai Institutes for Advanced Study, the University of Tokyo, Kashiwa, Japan 277-8583 (Kavli IPMU, WPI)}

\author{A.~Renzini}
\affiliation{Instituto Nazionale de Astrofisica, Osservatorio Astronomico di Padova, v.co dell'Osservatorio 5, I-35122, Padova, Italy, EU}

\author{C.~Mancini}
\affiliation{Dipartimento di Fisica e Astronomia, Universita di Padova, vicolo Osservatorio, 3, 35122, Padova, Italy}
\affiliation{Instituto Nazionale de Astrofisica, Osservatorio Astronomico di Padova, v.co dell'Osservatorio 5, I-35122, Padova, Italy, EU}

\author{F.~Bournaud}
\affiliation{Laboratoire AIM, CEA/DSM-CNRS-Universite Paris Diderot, Irfu/Service d'Astrophysique, CEA Saclay}

\author{A.~Puglisi}
\affiliation{Laboratoire AIM, CEA/DSM-CNRS-Universite Paris Diderot, Irfu/Service d'Astrophysique, CEA Saclay}
\affiliation{Instituto Nazionale de Astrofisica, Osservatorio Astronomico di Padova, v.co dell'Osservatorio 5, I-35122, Padova, Italy, EU}

\author{G.~Rodighiero}
\affiliation{Dipartimento di Fisica e Astronomia, Universita di Padova, vicolo Osservatorio, 3, 35122, Padova, Italy}

\author{D.~Liu}
\affiliation{Laboratoire AIM, CEA/DSM-CNRS-Universite Paris Diderot, Irfu/Service d'Astrophysique, CEA Saclay}

\author{M.~Sargent}
\affiliation{Astronomy Centre, Department of Physics and Astronomy, University of Sussex, Brighton, BN1 9QH, UK}

\author{N. Arimoto}
\affiliation{Astronomy Program, Department of Physics and Astronomy, Seoul National University, 599 Gwanak-ro, Gwanak-gu, Seoul, 151-742, Korea}
\affiliation{Subaru Telescope, 650 North A'ohoku Place, Hilo, Hawaii, 96720, USA}

\author{M. B\'{e}thermin}
\affiliation{European Southern Observatory, Karl-Schwarzschild-Str. 2, 85748 Garching, Germany}

\author{J.~Fensch}
\affiliation{European Southern Observatory, Karl-Schwarzschild-Str. 2, 85748, Garching bei M\"unchen, Germany}

\author{C.~C. Hayward }
\affiliation{Center for Computational Astrophysics, Flatiron Institute, 162 Fifth Avenue, New York, NY 10010, USA}

\author{J.~Kartaltepe}
\affiliation{School of Physics and Astronomy, Rochester Institute of Technology, 84 Lomb Memorial Drive, Rochester, NY 14623, USA}

\author{D.~Kashino}
\affiliation{Institute of Astronomy, ETH Z\"{u}rich, Wolfgang-Pauli-Strasse 27, CH-8093, Z\"{u}rich, Switzerland}

\author{A.~Koekemoer}
\affiliation{Space Telescope Science Institute, 3700 San Martin Drive, Baltimore, MD, 21218, USA}

\author{G. Magdis}
\affiliation{Cosmic DAWN Centre, Niels Bohr Institute, University of Copenhagen, Juliane Mariesvej 30, 2100, Copenhagen, Denmark}

\author{H.J. McCracken}
\affiliation{Institut d'Astrophysique de Paris, CNRS, UMR 7095 and UPMC, 98bis boulevard Arago, F-75014 Paris, France}

\author{T.~Nagao}
\affiliation{Graduate School of Science and Engineering, Ehime University, 2-5 Bunkyo-cho, Matsuyama 790-8577, Japan}

\author{K. Sheth}
\affiliation{NASA Headquaters, 300 E. St SW, Washington DC 20546, USA}

\author{V.~Smol\v{c}i\'c}
\affiliation{University of Zagreb, Physics Department, Bijeni\v{c}ka cesta 32, 10002, Zagreb, Croatia}

\author{F.~Valentino}
\affiliation{Dark Cosmology Centre, Niels Bohr Institute, University of Copenhagen, Juliane Maries Vej 30, DK-2100 Copenhagen, Denmark}



\begin{abstract}

We report on the discovery of a merger-driven starburst at $z=1.52$, PACS-787, based on high signal-to-noise ALMA observations. CO(5-4) and continuum emission (850$\mu$m) at a spatial resolution of 0.3$\arcsec$ reveal two compact ($r_{1/2}\sim1$ kpc) and interacting molecular gas disks at a separation of 8.6 kpc thus indicative of an early stage in a merger. With a SFR of 991 M$_{\odot}$ yr$^{-1}$, this starburst event should occur closer to final coalescence, as usually seen in hydrodynamical simulations. From the CO size, inclination, and velocity profile for both disks, the dynamical mass is calculated through a novel method that incorporates a calibration using simulations of galaxy mergers. Based on the dynamical mass, we measure (1) the molecular gas mass, independent from the CO luminosity, (2) the ratio of the total gas mass and the CO(1 - 0) luminosity ($\alpha_{CO}\equiv M_{gas}/L_{CO~1-0}^{\prime}$), and (3) the gas-to-dust ratio, with the latter two being lower than typically assumed. We find that the high star formation, triggered in both galaxies, is caused by a set of optimal conditions: a high gas mass/fraction, a short depletion time ($\tau_{depl}=85$ and 67 Myrs) to convert gas into stars, and the interaction of likely counter-rotating molecular disks that may accelerate the loss of angular momentum. The state of interaction is further established by the detection of diffuse CO and continuum emission, tidal debris that bridges the two nuclei and is associated with stellar emission seen by HST/WFC3. This observation demonstrates the power of ALMA to study the dynamics of galaxy mergers at high redshift.

\end{abstract}





\section{Introduction}

\begin{figure*}
\epsscale{1.0}
\plotone{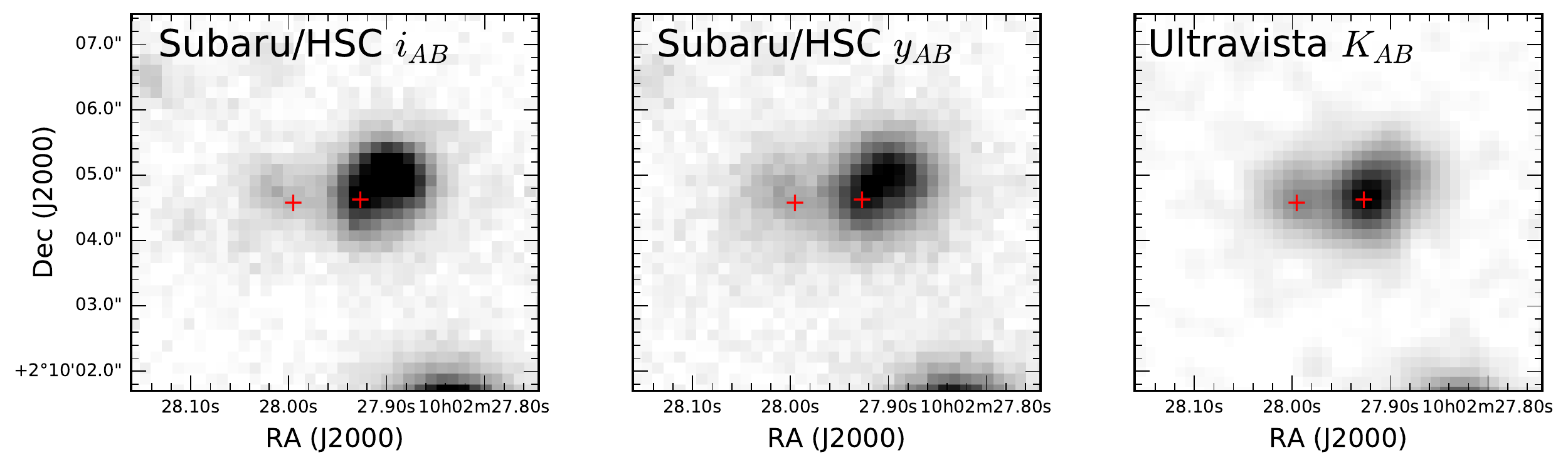}
\epsscale{0.8}
\plotone{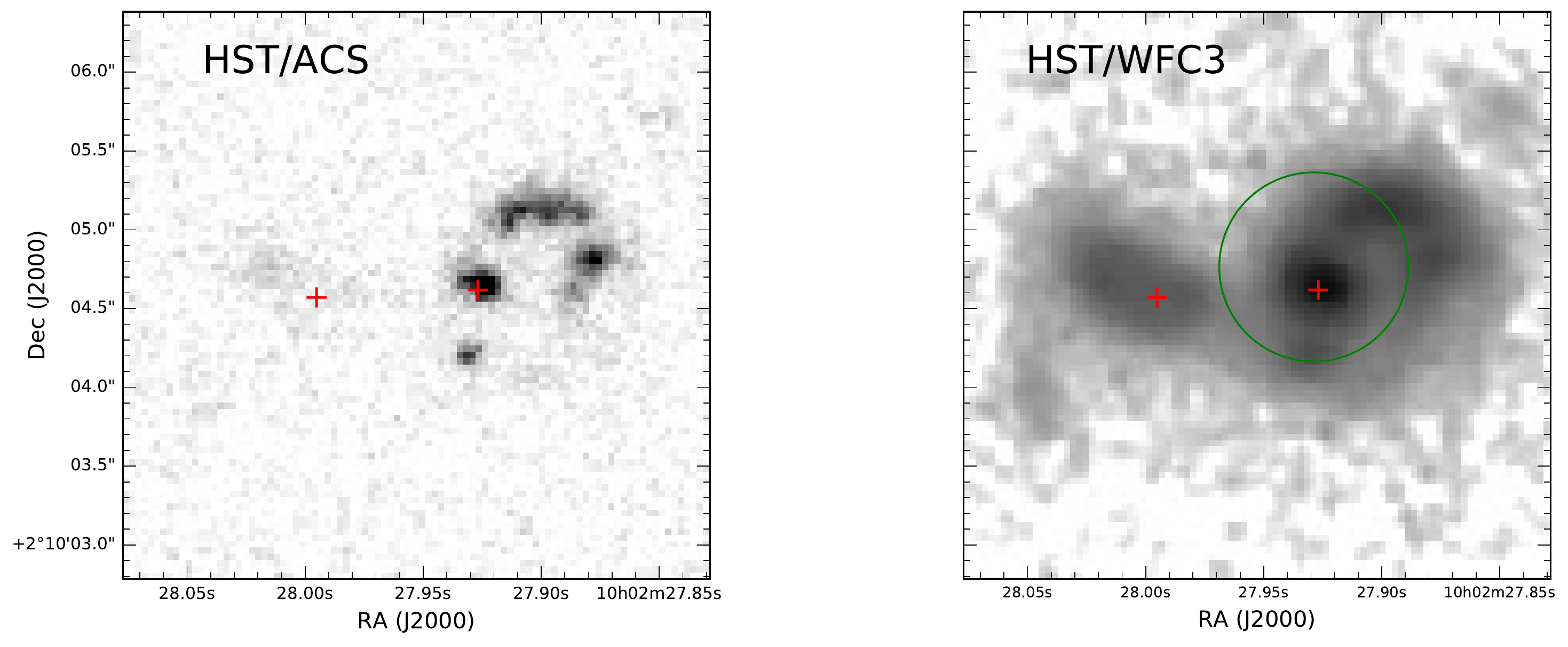}
\caption{Optical-to-infrared imaging of PACS-787. $Top~row~(ground-based):$ The two left panels show the Subaru/HSC optical images while the UltraVISTA \citep{Mccracken2012} near-infrared image is on the right. The two galaxies are visible in each frame. A small red cross indicates the centroid of each component based on the CO(5 - 4) emission that is co-spatial with the IR but offset with the locations in the bluer images, as typically seen for high-z starbursts \citep{Silverman2015a} and more typical high-z galaxies \citep{Cibinel2017}. $Bottom~row~(space-based):$ HST imaging in the ACS F814W and WFC3 F140W filters with a linear and log scale respectively. The green circle indicates the location of the fiber used to acquire a NIR spectrum with Subaru/FMOS.}
\label{fig:images}
\end{figure*}

Our understanding of the physical properties of the interstellar medium, in galaxies undergoing a merger that can induce star formation and accretion onto supermassive black holes, is usually based on studies in the local Universe. Representative examples include Arp 220, the Antennae, and NGC 6240 that belong to a class of Ultraluminous Infrared Galaxies (ULIRGs; \citealt{Sanders1996}) that are unequivocally undergoing an extreme starburst (with a luminous Active Galactic Nucleus - AGN - in many cases), triggered by a merger of two massive galaxies \citep[e.g.,][]{DiMatteo2005,Hopkins2006,Volonteri2015}. While such collisions accelerate the growth of galaxies \citep[e.g.,][]{Scudder2012,Ellison2013}, this mechanism appears to have a subdominant effect on the buildup of the global stellar mass and black hole mass density over a wide range of cosmic time \citep{Rodighiero2011}. Although, mergers likely represent a key phase (having a short duty cycle) through which every massive galaxy may pass \citep[e.g.,][]{Lotz2008,Lackner2014}. Moreover, galaxy mergers offer us a specialized laboratory to study the physics of the interstellar medium including the conversion of molecular gas into stars and the channeling of such gas to the central regions that can build up the central stellar bulge.

From seminal studies, local starbursts are characterized as having relatively large amounts of molecular gas \citep{Sanders1986,Scoville1989}, hence higher gas fractions relative to the normal galaxy population at $z\sim0$. The majority of the molecular gas is centrally concentrated within a radius of $\sim1$ kpc \citep{Downes1998,Wilson2008}, possibly the result of a nuclear inflow induced by the merger. This has been demonstrated to be a common phenomenon through hydrodynamic simulations \citep[e.g.,][]{Hopkins2013,Fensch2017} of the merger of two gas-rich disk galaxies. Furthermore, starbursts are observed to have a lower gas content than expected given their high star formation rates (SFRs) that has been described as a higher efficiency ($SFE\equiv SFR/M_{gas}$) to form stars \citep{Solomon1997,Saintonge2012}, even seen to be present at higher redshift \citep{Daddi2010b, Genzel2010,Combes2013,Tacconi2018}.

Numerous millimeter interferometric studies have presented observations of the molecular gas and dust content of starburst galaxies at high redshift ($z>1$), usually bright submillimeter galaxies \citep[e.g.,][]{Greve2005,Tacconi2006,Tacconi2008,Engel2010,Bothwell2013}. For select starbursts, lensing has enhanced the ability to resolve the emission on subarcsecond scales \citep[e.g.,][]{Swinbank2010,Spilker2015}. In general, submillimeter galaxies share many properties with local ULIRGs such as having large amounts of molecular gas, clear signs of mergers in many cases (and ordered disk rotation), short gas depletion times, and characteristic sizes less than a few kiloparsecs with the central regions being baryon dominated. These galaxies have been proposed to be the progenitors of local elliptical galaxies \citep{Kormendy1992,Ivison1998,Lilly1999,Genzel2001} with an intermediate stage as a compact quiescent galaxy \citep{Toft2014}. Although, a non-negligible level of contamination from unassociated galaxies is present \citep{Hayward2013} due to the large beam size of submillimeter-selected sources that is now being mitigated with Atacama Large Millimeter/submillimeter Array (ALMA).  

With ALMA, many efforts are extending the aforementioned studies of the molecular gas and dust emission in galaxy mergers and starbursts at $z>1$, particularly below the kiloparsec scale due to the large collecting area and long baselines \citep[e.g.,][]{Chen2017, Barro2017,Riechers2017}. For example, \citet{Hodge2016} has measured the effective radius of submillimeter galaxies ($z\sim2.5$), based on the size of the dust emission at 870$\mu$m, to be between 1 - 3 kpc and have a disk-like morphology, in contrast to star formation being more extended on galaxy-wide scales for typical star-forming main sequence (MS hereafter) galaxies at $z\sim2$ \citep{Rujopakarn2016}. While there are signs of ongoing mergers in HST images of these SMGs, the dust emission is slightly more extended than local ULIRGs while having similar L$_{\rm IR}$ (10$^{12-13}$ L$_{\odot}$). These results corroborate evidence for central $\sim$kpc dust-emitting (i.e., star-forming) regions in high-z SMGs \citep{Simpson2015} that may indicate a phase where massive galaxies are rapidly forming their bulges given their high star formation rate surface density ($\Sigma _{SFR}\sim 100$ M$_{\odot}$ yr$^{-1}$ kpc$^{-2}$). However, these issues are still under debate and will benefit from larger samples, observed at higher resolution with ALMA, including cases (e.g., PACS-787) with more extreme rates of star formation as presented here.

Complementary to these studies, we are carrying out a program with ALMA over multiple cycles to measure the molecular gas properties, primarily using carbon monoxide (CO; 2 - 1), of twelve starburst galaxies in the COSMOS field having star formation rates well-above ($\gtrsim4\times$) the star-forming MS at $z\sim1.6$ \citep[see][]{Silverman2015a,Silverman2018}. This sample has advantages over other heterogeneous high-z starburst samples by having a wealth of information including accurate spectroscopic redshifts, metallicities from [NII]/H$\alpha$ \citep{Puglisi2017}, and stellar masses and SFRs based on mid- and far-IR observations with $Spitzer$ \citep{Sanders2007} and $Herschel$ \citep{Lutz2011}, placing them securely in the starburst class \citep{Rodighiero2011}. Compared to classical submillimeter galaxies (SMG), this selection avoids contamination by objects close to the MS \citep{Magnelli2012}, reduces SED bias compared to SMGs selection (which prefers cold objects), and is able to pick objects even with relatively moderate SFR,  but much higher above the MS. Based on this sample, we demonstrate that the star formation efficiency (or gas depletion timescale) increases (decrease) continuously with elevation in SFR above the star-forming MS \citep{Silverman2015a,Silverman2018}.

Here, we present a study at higher spatial resolution of the CO(5-4) molecular gas and continuum emission (rest-frame 500$\mu$m) of our strongest starburst (PACS-787; $z_{spec}=1.5234$; SFR = $991_{-87}^{+96}$ M$_{\odot}$ yr$^{-1}$, 6.5$\times$ greater than the average star-forming galaxy for its stellar mass) from the sample described above, using ALMA in Cycle 4 \citep{Silverman2018}. With $0.3\arcsec$ resolution in Band 6, we cleanly separate the emission from two separate galaxies in a major merger. With the resulting maps (CO and continuum) having a high signal-to-noise, we assess the physical characteristics of each galaxy through a determination of $\alpha_{CO}$ (and the gas-to-dust ratio) based on a dynamical mass estimate of PACS-787. As a novelty, we apply a scale factor to calculate the dynamical mass based on hydrodynamical simulations of galaxy mergers with similar properties to PACS-787. Subsequently, we determine gas mass (M$_{gas}$), SFR, and star formation efficiency (M$_{gas}$/SFR) of each galaxy. To conclude, we argue that PACS-787 is a classic example of a merger-induced starburst with a number of contributing factors to its SFR including an orbital configuration conducive for forming stars at such a high rate. Throughout this work, we assume an established cosmology ($H_0=70 $ km s$^{-1}$ Mpc$^{-1}$, $\Omega_{\Lambda}=0.7$, $\Omega_{\rm{M}}=0.3$), and use a Chabrier IMF for SFRs and stellar masses.





\section{PACS-787: an extreme outlier from the SF MS}
 
PACS-787 is a starburst galaxy with a high SFR as compared to the typical star-forming population at fixed redshift and stellar mass (with details on the determination of these measurements to follow). The galaxy was first identified as having strong far-infrared emission, as detected by $Herschel$/PACS \citep{Rodighiero2011}, and a photometric redshift within a targeted redshift range of $1.43 < z < 1.74$ to detect the H$\alpha$ emission line as part of our large near-infrared spectroscopic survey of star-forming galaxies, FMOS-COSMOS \citep{Silverman2015b}. Therefore, the object was given priority for spectroscopic follow up that resulted in a positive detection of both H$\alpha$ and [NII]$\lambda$6584 at z=1.5234 (Appendix~\ref{sec:AGN}) thus confirming its nature as having a very high SFR, atypical of star-forming MS galaxies. 

In Figure~\ref{fig:images}, we show image cutouts of PACS-787 in the optical and near-infrared from both the ground and space that indicate the presence of two galaxies in close proximity. While the eastern (left-most) component is faint in the optical, it is clearly detected by Subaru/Hyper Suprime-Cam (HSC) and UltraVISTA (K$_{AB}$=22.39 and 20.47 for the East and West galaxies respectively). High spatial resolution HST imaging with ACS/F814W \citep{Koekemoer2007, Scoville2007} and WFC3/F140W (single-orbit; 40 minute exposure on-source), with the latter provided through an unrelated Cycle 25 program (PI J. Silverman; Program \#15115), aids in our characterization of the system as an ongoing merger. For instance, the two galaxies are undergoing a strong interaction as indicated by the presence of tidal features and a stellar bridge between the two nuclei seen in the WFC3 image. The HST/ACS mosaics and the WFC3 image (Koekemoer 2018, priv. comm) have been reprocessed where the astrometry is directly tied to the International Celestial Reference Frame (ICRF), thereby eliminating uncertainties from possible offsets in older catalogs previously used to align HST images. For ACS, this correction resulted in a 5 pixel ($0.15\arcsec$) shift. The new astrometric solution for both images shows near perfect alignment between the bright UV and IR emission of the western galaxy with the ALMA centroid of the CO(5-4) emission as shown in the bottom panels of Figure~\ref{fig:images} by the red crosses. Here, we provide details on the use of the multi-wavelength photometry available from the COSMOS survey to accurately measure the stellar mass and SFR, both key quantities required in the analysis of the ALMA observations, of the individual galaxies.

\begin{figure}
\epsscale{1.2}
\plotone{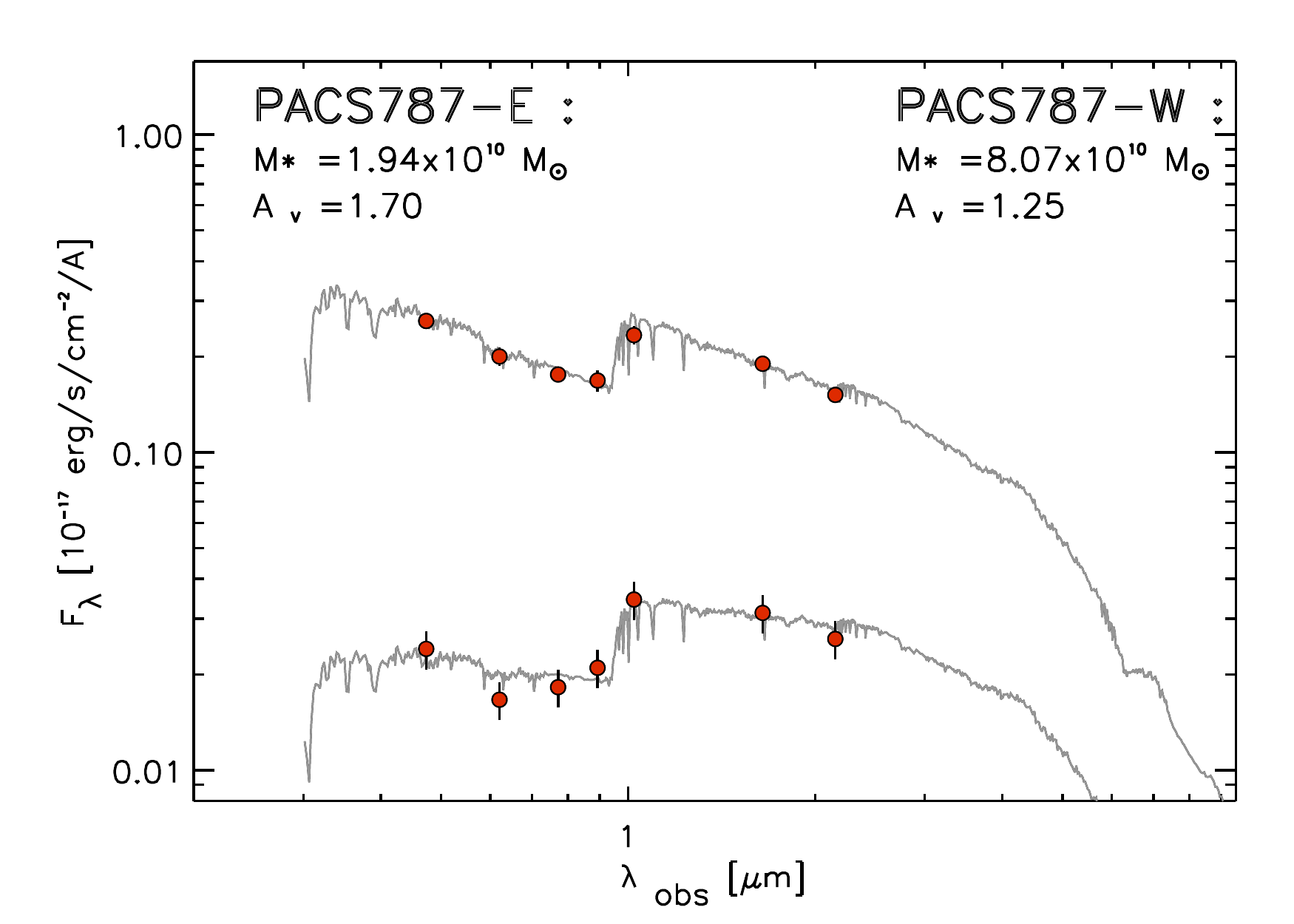}
\caption{Best-fit stellar population model to optical-infrared photometry for the eastern (fainter) and western (brighter) galaxies separately with the respective best-fit stellar masses and extinction A$_{V}$ as labeled.}
\label{fig:sed_1}
\end{figure}

\subsection{Stellar mass ($M_{stellar}$)}
\label{text:mass}

The mass, in stars, that have already formed is a key measurement for insight into the growth history of galaxies and the current mode of star formation such as being in a starburst phase, as is the case for PACS-787. Multiple broad-band images of the COSMOS field, covering the ultraviolet to the infrared, are used to measure the stellar mass as done in \citet{Laigle2016}. Since the galaxies are slightly blended, we use GALFIT \citep{Peng2010} to measure fluxes in seven optical bands (g, r, i, z; Subaru/HSC; \citealt{Tanaka2017} and near-infrared (Y, H, Ks; UltraVISTA; \citealt{Mccracken2012}), for each component separately. To measure the stellar mass, we fit the photometry using models \citep{Bruzual2003} that include a mixture of different stellar populations. The code Hyperzmass \citep[i.e., a modified version of Hyperz,][]{Bolzonella2000} is used to perform the fit while implementing a constant star formation history (SFH). The stellar mass estimate accounts for stellar mass loss as described in \citet{Pozzetti2007}. We highlight that degeneracies between parameters such as stellar age, metallicity, and extinction affect the stellar mass estimates and are accounted for in the 68\% confidence interval based on the $\chi^2$ statistic.

In Figure~\ref{fig:sed_1}, we show the best-fit model to the spectral-energy distribution for the separate components of PACS-787. We note that the assumption of a constant SFH will result in lower stellar masses than if assuming an exponentially declining model with a recent burst. However, the sum of the two best-fit SEDs gives mid-IR fluxes which are within 10\% of the measured {\it Spitzer}/IRAC fluxes, indicating that the two galaxies do not contain additional mass that is unaccounted for by the photometry used for the SED fit. Continuum emission from the type 2 AGN (Appendix~\ref{sec:AGN}) in the western galaxy is negligible thus does not impact the stellar mass measurements. The stellar masses of each, with uncertainties based on photometric errors, are reported in Table~\ref{tab:source_prop} and shown in Figure~\ref{fig:sed_1} assuming a Chabrier IMF. The galaxies have a stellar mass ratio of 4.2:1 thus indicative of a marginal major merger. However, when also considering the mass of the molecular gas (see Section~\ref{co_mass}) the mass ratio becomes $\sim 2:1$, which definitely qualifies the system as a major merger.

\begin{figure}
\epsscale{1.2}
\plotone{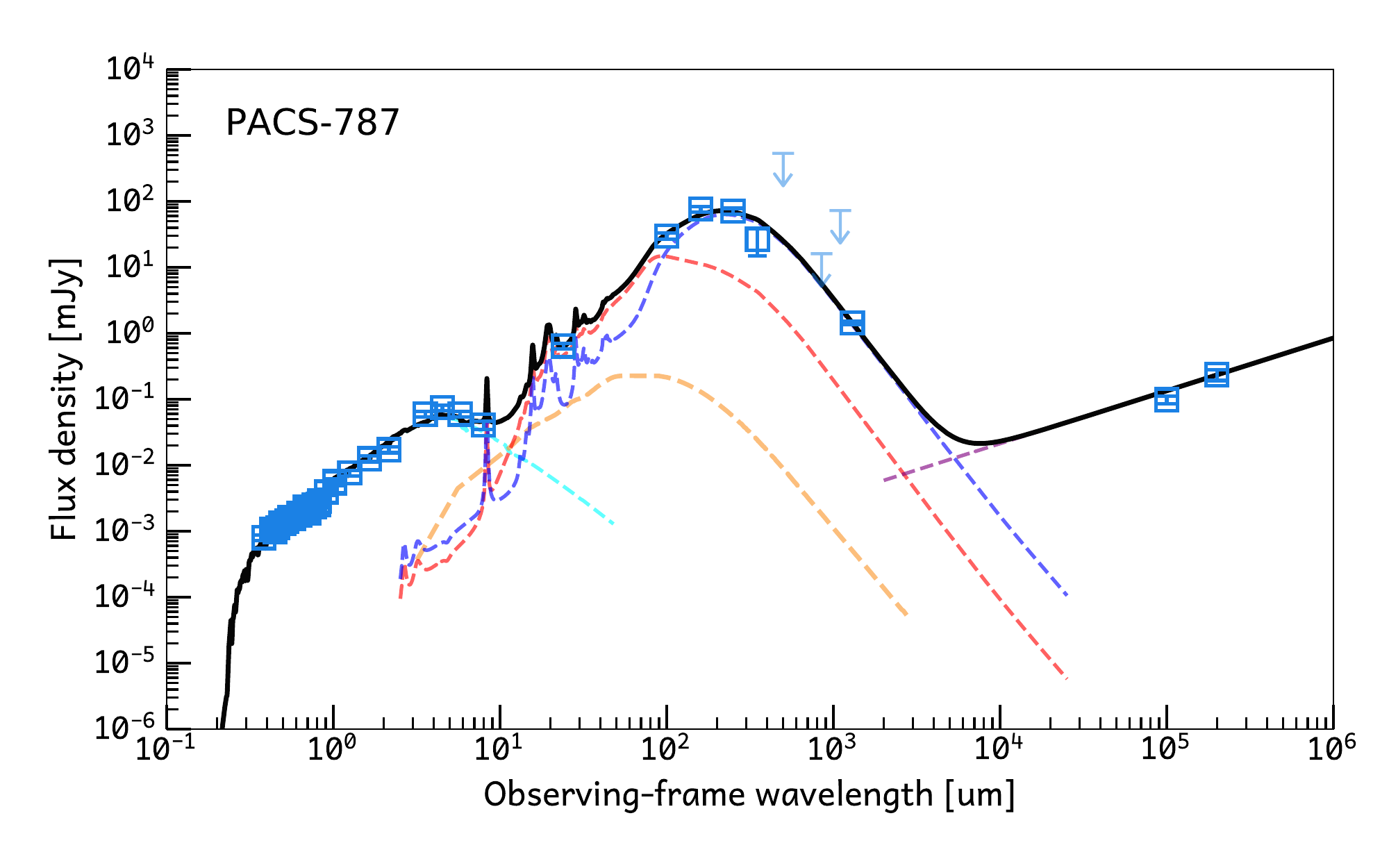}
\caption{SED of PACS-787 (Total; East + West) with best-fit dust model \citep{Draine2007} to the FIR emission. The individual components to the fit are shown in color (stars: cyan; AGN - yellow; warm dust - red; cool dust - purple). Blue squares show photometric measurements and their 1$\sigma$ errors. A power law radio spectrum highlights the agreement between star formation rate indicators.}
\label{fig:sed_2}
\end{figure}

\subsection{Star formation rate (SFR)}
\label{text:sfr}

Far-infrared (FIR) emission, primarily from cool and warm dust heated by young massive stars, has been demonstrated to be an accurate indicator of the SFR of galaxies \citep{Kennicutt2012}, particularly starbursts which are heavily enshrouded by dust. Hence, we measure the total FIR luminosity that is determined from an integral of a best-fit dust model over the wavelength range 8 - 1000 $\mu$m.

Dust emission is measured using models \citep{Draine2007} that fit the observations covering the mid- to far-infrared wavelengths \citep[e.g.,][]{Magdis2012a}. As shown in Figure~\ref{fig:sed_2}, continuum emission from PACS-787 is detected by Spitzer/MIPS (24 $\mu$m), Herschel/PACS (100 and 160 $\mu$m), Herschel/SPIRE (250, 350 and 500 $\mu$m) and ALMA (1.28 mm) that covers the extent of the FIR spectral-energy distribution (SED). Following \citet{Magdis2012a}, we simultaneously fit the broad-band SED using a model with five components: stellar light, warm and cool dust, an AGN, and radio power-law emission. The latter is decoupled from the global fit. We do not require energy balance between the FIR and absorbed UV emission since the dust and UV radiation in starbursts are not co-spatial as the case for PACS-787 and our full starburst sample \citep{Silverman2015a,Puglisi2017}.

The dust model is expressed by Equation 2 of \citet{Magdis2012a} with seven free parameters ($\Omega_{*}$, $q_{PAH}$, $U_{min}$, $U_{max}$, $\alpha$, $\gamma$, $M_{dust}$). The stellar component is approximated by a blackbody ($T_{*}=5000$ K) with $\Omega_{*}$ being the solid angle subtended by the stellar photospheres. The fraction of the dust in PAH grains is expressed as $q_{PAH}$. The radiation field is given as $U_{min}$ that illuminates most of the diffuse dust while a smaller fraction ($\gamma$) is exposed to a higher radiation field up to a value of $10^6$ for $U_{max}$. An index $\alpha$, fixed at a value of 2, describes the distribution of the starlight intensity incident on the dust (d$M_{dust}$/d$U\propto U^{-\alpha}$). The best-fit model is found through a $\chi^2$ minimisation with a goodness-of-fit given by the reduced $\chi^2 \def \chi^2/N_{\rm dof}$. From this method, we find a total dust mass of $log~M_{dust}=9.2\pm{0.06}$ M$_{\odot}$ based on the fit shown in Figure~\ref{fig:sed_2}. Monte Carlo simulations are used to assess the error on $M_{dust}$.

From an integral of the best-fit dust model, the total FIR luminosity is $log~L=12.83\pm0.04$ (units of $L_{\odot}$) that corresponds to a total SFR = $991_{-87}^{+96}$ M$_{\odot}$ yr$^{-1}$, using the calibration given in \citet{Kennicutt2012}, which includes the contribution from both galaxies. This rate is in good agreement with that derived from the radio emission (3 and 5 GHz) \citep{Schinnerer2007,Smolcic2017} where the normalization of the power-law model, shown in Figure~\ref{fig:sed_2}, is set by the IR-radio correlation with a parameterization from \citet{Ivison2010}.

\begin{figure}
\epsscale{1.2}
\plotone{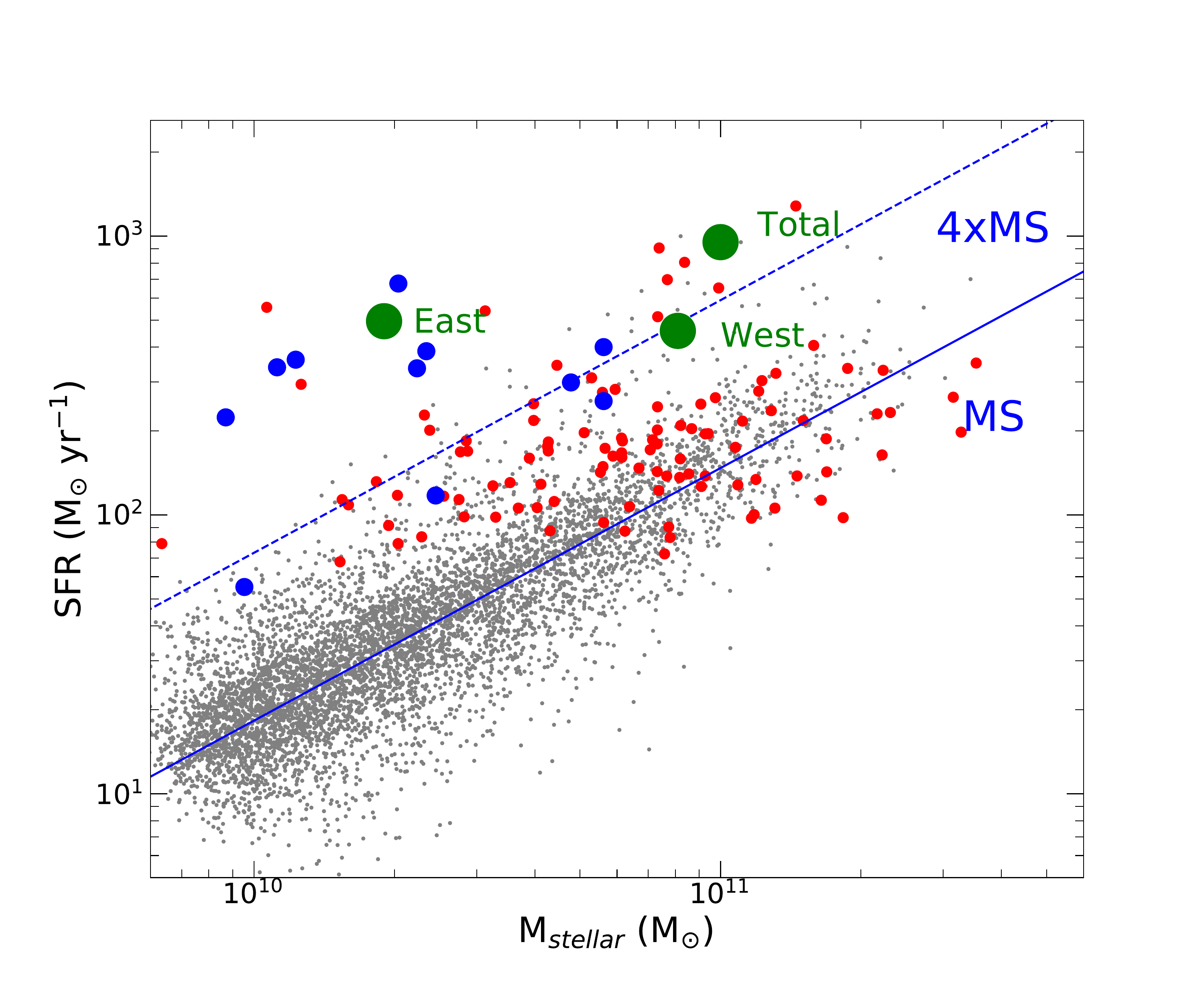}
\caption{Location of PACS-787 on the star-forming MS (SFR versus M$_{stellar}$). Both the sum (total) and the individual (East and West) galaxies are indicated by large green circles with the latter based on high-resolution ALMA continuum observations (Section~\ref{text:split_sfr}). Galaxies with $1.4 < z < 1.7$ in COSMOS are shown for comparison (MS galaxies - grey dots; red symbols - $Herschel$-detected). The blue circles indicate those having CO(2 - 1) observations with ALMA. Slanted lines indicate the MS (solid; z = 1.5) and a parallel track at an elevated rate of 4$\times$ (dashed).}
\label{fig:MS}
\end{figure}

\subsection{Location relative to the star-forming MS}

In Figure~\ref{fig:MS}, we display the location of PACS-787 with respect to the star-forming galaxy population at $1.4<z<1.7$ that delineates the known star-forming MS \citep[e.g.,][]{Speagle2014,Renzini2015}. For this comparison, we measure the stellar masses consistently between both samples to determine, as accurately as possible, the boost in SFR for PACS-787, at a given mass, that is used subsequently in our analysis. Star-forming galaxies are selected from the COSMOS photometric catalog \citep{Laigle2016} with $1.4<z <1.7$. Stellar masses for this reference sample are remeasured, as described in \citet{Puglisi2017}, using the same code, and assumptions on the equivalent measurements for PACS-787. The only difference is that for PACS-787 we use the latest Subaru/HSC data. For the comparison sample, SFRs are based on the best-fit stellar population synthesis model. Reassuringly, the locus of the MS agrees very well with published relations as given in \citet{Daddi2007} and \citet{Kashino2013}. For the $Herschel$-detected galaxies (shown in red), the SFRs are computed based on the FIR-SFR relation \citep{Kennicutt2012}. Considering here the sum of the stellar mass for both galaxies, we show the location of PACS-787 at an elevation $6.5\times$ higher with respect to the star-forming MS, thus securely undergoing a starburst event. 

\section{ALMA observations and data analysis}

\begin{figure}
\epsscale{1}
\plotone{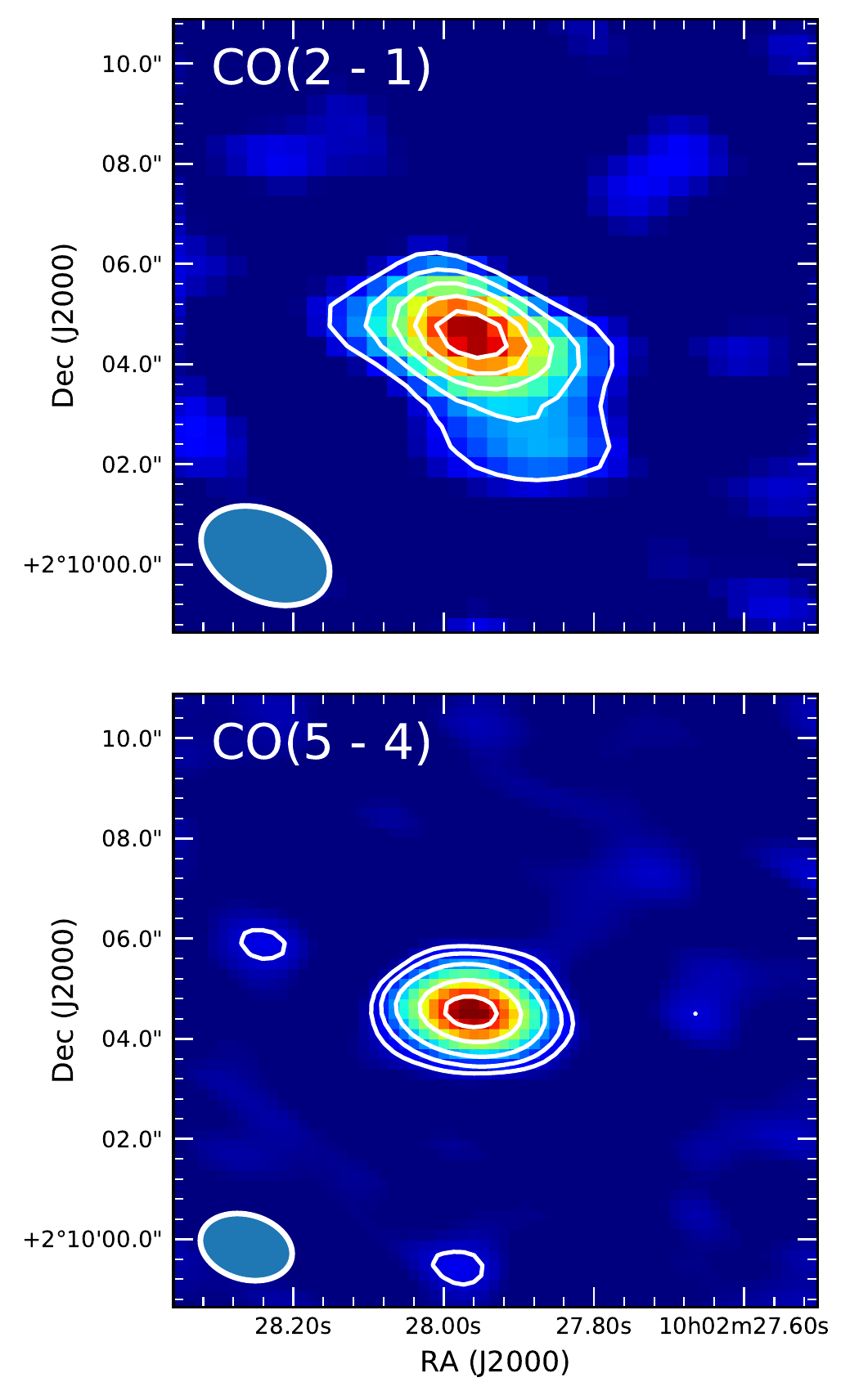}
\caption{Low resolution ALMA imaging of PACS-787. (top) CO(2 - 1) map with contours at 2, 4, 6, 8, $10\times \sigma_{rms}$ where $\sigma_{rms}=0.085$ Jy beam$^{-1}$ km s$^{-1}$.   (Bottom) CO(5 - 4) map with contours at 2, 4, 8, 16, $24\times \sigma_{rms}$ where $\sigma_{rms}=0.1487$ Jy beam$^{-1}$ km s$^{-1}$. The ALMA beamsize is shown in the lower left of each panel.}
\label{fig:lowres}
\end{figure}

As part of our program to measure the molecular gas properties of high-z starbursts \citep{Silverman2015a,Silverman2018}, we obtained a low resolution (1.7$\arcsec$) observation of PACS-787 in ALMA Cycle 3 (Project 2015.1.00861.S) to detect CO(2 - 1) at 91.31 GHz (Band 3). This target has the brightest CO emission of our full sample with $I_{CO [2-1]}=1.64\pm0.20$ Jy km s$^{-1}$. The emission, reproduced in Figure~\ref{fig:lowres}, is indicative of a single source with marginal evidence of being extended. As shown in Figure 5 of \citet{Silverman2018}, there are signs of different kinematic components. The CO(2-1) luminosity of PACS-787, based on this observation, is used in this study (Section~\ref{text:aco}) to measure the ratio of $M_{gas}$ to $L_{CO~1 - 0}^{\prime}$ ($\alpha_{CO}$) with $M_{gas}$ being assessed independent of the CO luminosity.

To determine the nature of this extreme starburst, we acquired a higher resolution CO(5 - 4) map of PACS-787, along with two additional starburst galaxies (PACS-830 and PACS-819) elevated above the MS star-forming population at similar redshifts. Here, we report on the observations of PACS-787 by itself since the nature of the source is distinct from the others with clear evidence for two gas-rich interacting galaxies at an early stage of the merger. In Table~\ref{tab:alma_obs}, we list the ALMA observations used in this study.

\begin{deluxetable*}{llllllccc}
\tabletypesize{\scriptsize}
\tablecaption{Summary of ALMA observations used in this study\label{tab:alma_obs}}
\tablehead{\colhead{Obs. date}&\colhead{Band}&\colhead{Spectral}&\colhead{Central frequency}&\colhead{Band width}&\colhead{Channel width}&\colhead{Integration time}&\colhead{Beam size}&\colhead{$\sigma_{rms}$}\\
\colhead{(UT)}&&\colhead{feature}&\colhead{(GHz)}&\colhead{(GHz)}&\colhead{(MHz)}&\colhead{(min)}&\colhead{($\arcsec$)}&\colhead{(mJy beam$^{-1}$)}}
\startdata
2016Mar04&3&CO(2 - 1)&89.00, {\bf 91.00\tablenotemark{a}}&1.875, 1.875&1.95, 1.95&6.55&$2.7\times1.8$&0.085$\tablenotemark{b}$\\
&&&101.10,103.00&1.875, 1.875&15.6, 15.6\\
2016Nov07&6&CO(5 - 4)&226.01, {\bf 228.43\tablenotemark{a}}&2.0, 1.86&15.62, 3.91&19.73&$0.4\times0.3$&$0.045\tablenotemark{b}$\\
&&&241.01, 244.01&2.0, 2.0&15.62, 15.62&&\\
2017Mar18&6&CO(5 - 4)&226.01, {\bf 228.43\tablenotemark{a}}&2.0,2.0&15.62, 15.62&10.2&$1.9\times1.3$&$0.186\tablenotemark{b}$\\
&&&241.01, 244.01&2.0, 2.0&15.62, 15.62\\
\enddata
\tablenotetext{a}{The central frequency in bold indicates the spectral window for which the CO line falls within.}
\tablenotetext{b}{rms noise level for a channel width of 600 km s$^{-1}$ centered on the CO emission-line}
\end{deluxetable*}

\subsection{High resolution: CO(5 - 4) and continuum emission at 1.28 mm }

PACS-787 was observed on November 7, 2016 with 43 12m antennas as part of our Cycle 4 program (Proposal \#2016.1.01426.S; PI J. Silverman). The antenna baselines spanned a range of 18.6 m to 1.11 km that resulted in beam size of $0.38\arcsec \times0.34\arcsec$ and a maximum recoverable scale of 2.98$\arcsec$. Four spectral windows were assigned within Band 6 to detect the CO (5 - 4; $\nu_{rest}$= 576.27 GHz) emission line and underlying continuum. The on-source integration time was 19.73 minutes. The calibration targets are J1058+0133 (bandpass, flux, pointing), and J0948+0022 (phase).

As part of the same program, PACS-787 was re-observed on March 18, 2017 with 42 antennas in a compact configuration having a maximum baseline of 278.9 m thus resulting in a beam size is $1.9\arcsec \times 1.3\arcsec$. The purpose of this observation was to measure the total CO(5 - 4) emission including any component (see Section~\ref{sec:diffuse}) that may be undetected on larger scales with the higher resolution observations. We employed the same four spectral windows in Band 6 as for the high-resolution observations. The target was observed for 10.2 minutes (on-source). Observations of standard calibration sources were taken as mentioned above. In the bottom panel of Figure~\ref{fig:lowres}, we show the low resolution CO(5-4) image of PACS-787.

\subsection{Emission-line and continuum measurements}

We use Common Astronomy Software Applications (CASA) to carry out the processing of the ALMA raw data though the standard pipeline. From the visibility files, we then generate images using the task `clean' with a natural weighting scheme of the antenna pattern. The image cube is collapsed using `immoment' over a velocity range (given as $\Delta {\rm v}$ in Table~\ref{tab:source_prop}) that encompasses the CO(5 - 4) line. Continuum images make use of the full spectral range without channels having CO emission. We perform scientific analysis on the fits images as described below. Measurements are confirmed using a separate analysis package, GILDAS, and the task `uvmodelfit'.

\begin{figure*}
\epsscale{1.0}
\plotone{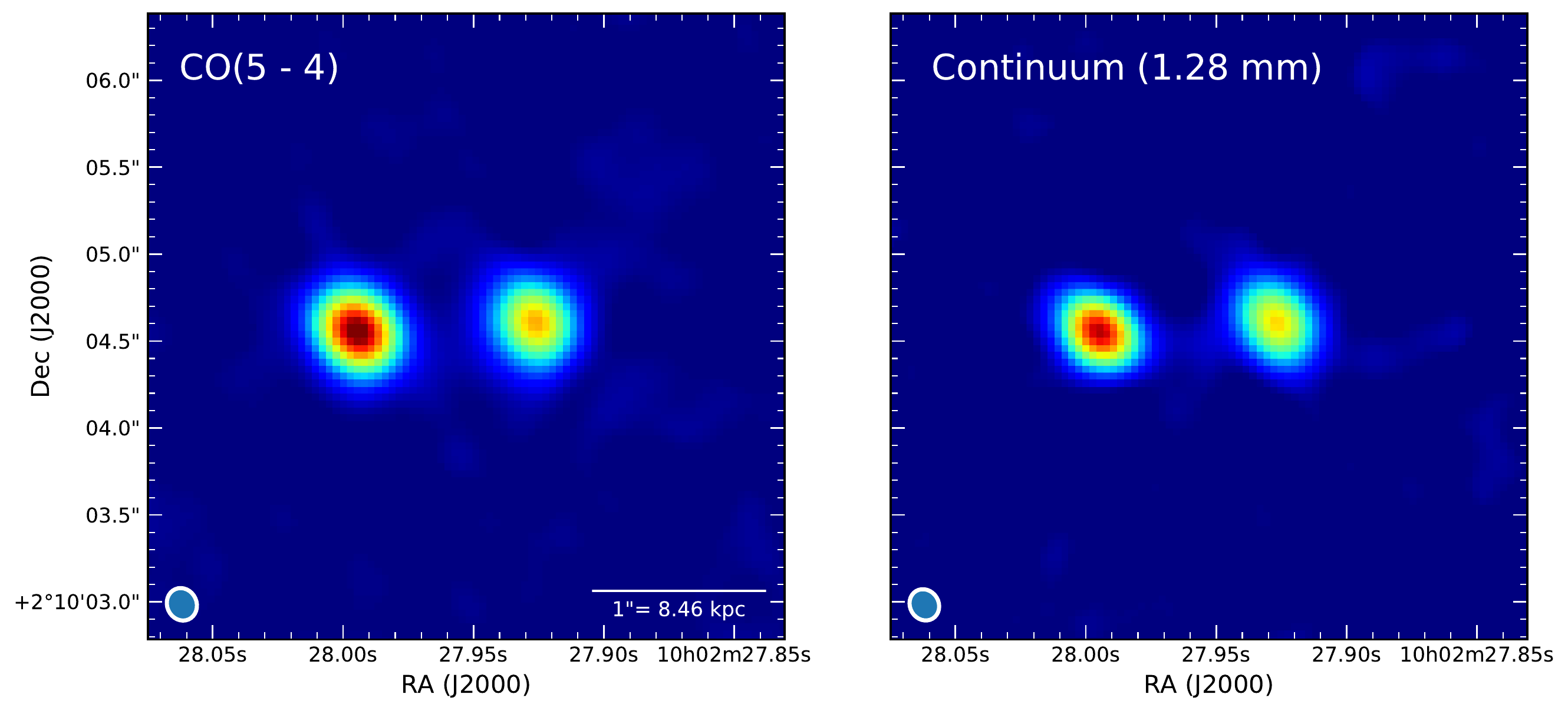}
\epsscale{1.0}
\plotone{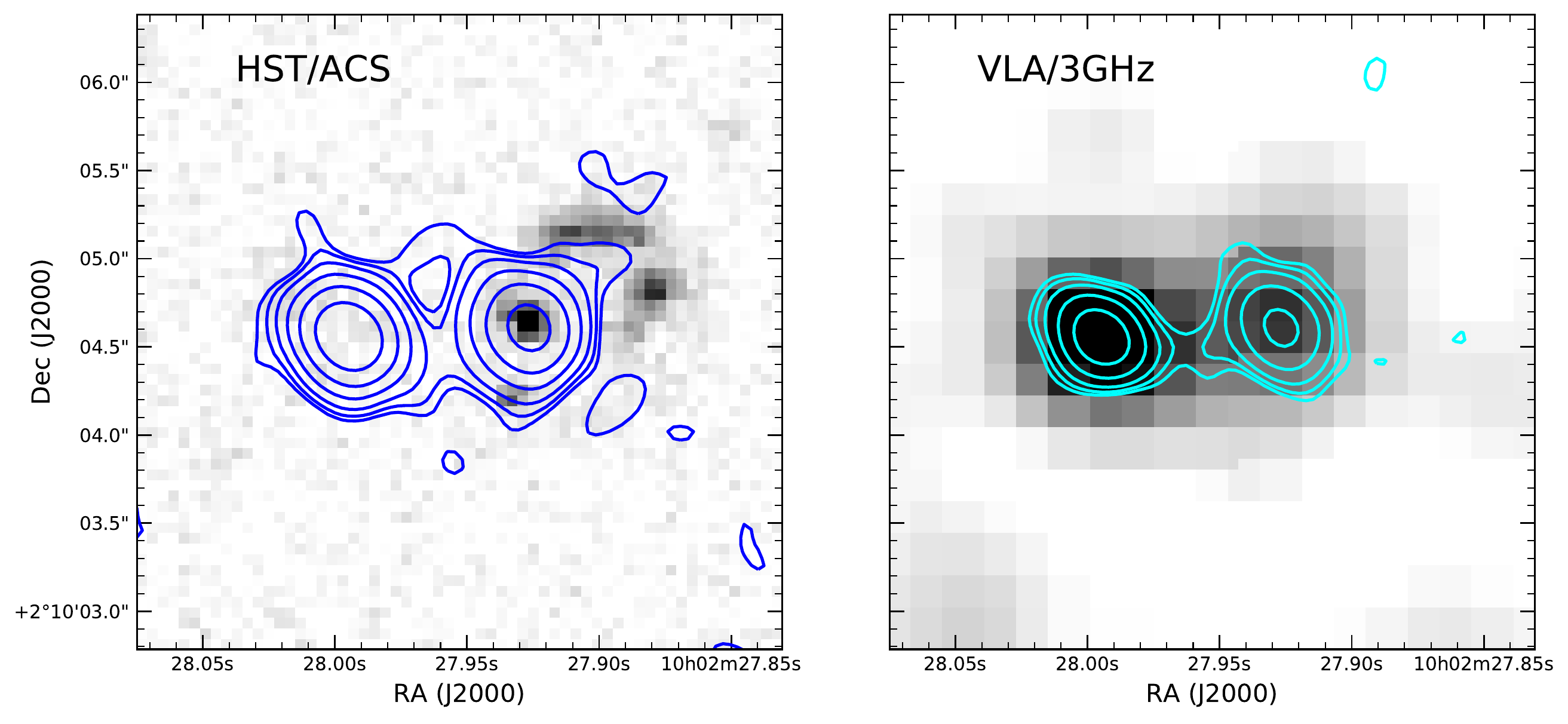} 
\caption{ALMA imaging of PACS-787 reveals two separate galaxies, both bright in CO (5 - 4; top-left panel) and continuum (1.28 mm; top-right panel), in the early stages of an interaction given the projected separation of 8.6 kpc. The ALMA beam size is shown in the lower left corner of the top two panels. HST/ACS imaging (bottom-left panel) only detects diffuse tidal features and a bright component in the Western galaxy. The blue contours indicate the CO brightness level at 2, 3, 5, 10, 20, $40\times \sigma_{rms}$ where $\sigma_{rms}=0.039$ Jy beam$^{-1}$ km s$^{-1}$. In the bottom right panel, the ALMA continuum emission (aqua blue contours at the same multiples of $\sigma_{rms}= 2.037\times 10^{-5}$ Jy beam$^{-1}$ as the left panel) is in excellent agreement with radio emission at 3 GHz \citep{Smolcic2017}, indicative of their ongoing star formation. In each panel, North is up and East is to the left.}
\label{fig:alma}
\end{figure*}

\section{Observational results}

\subsection{Molecular line emission (CO 5 - 4)}

\begin{figure}
\epsscale{1.8}
\plottwo{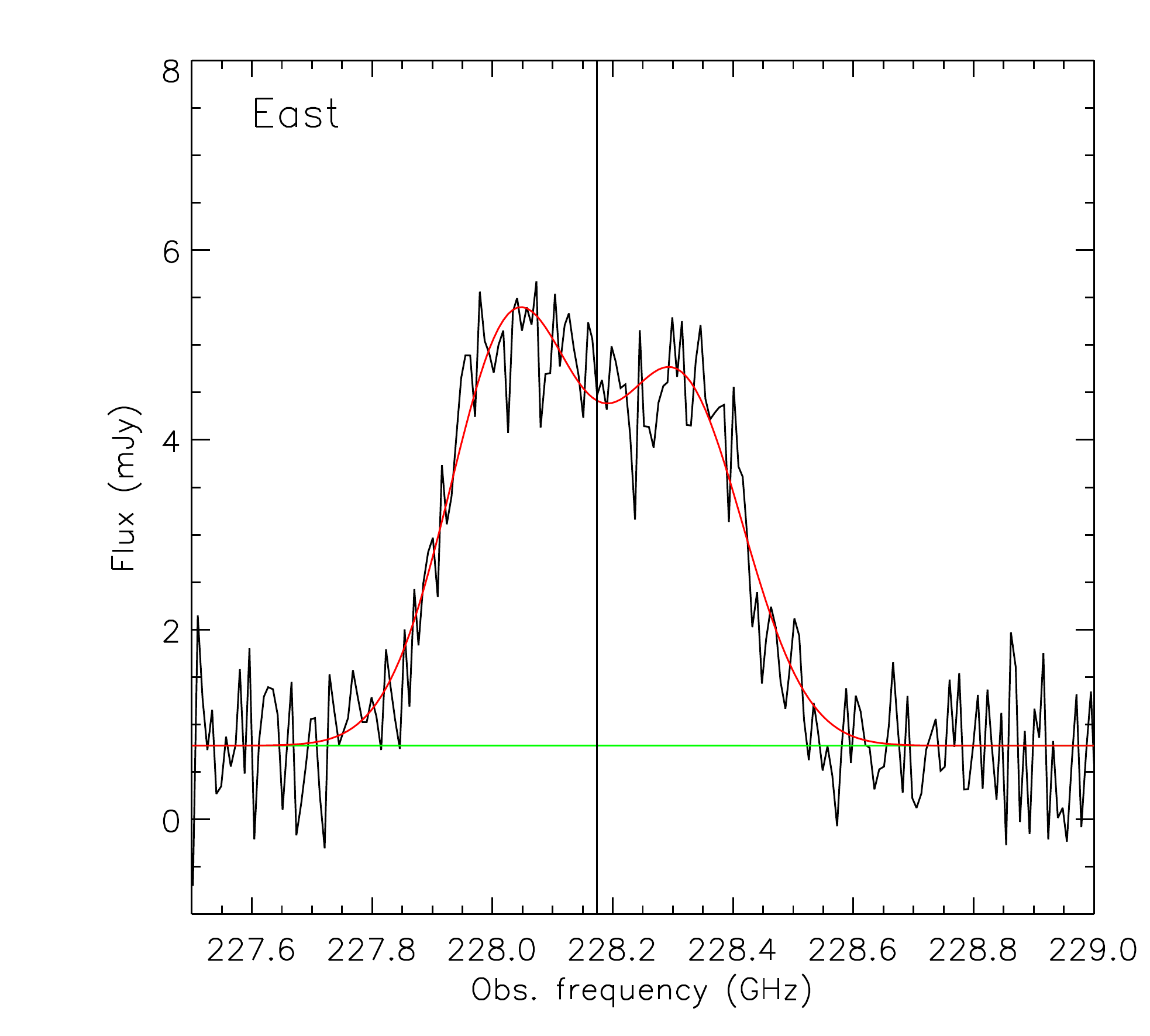}{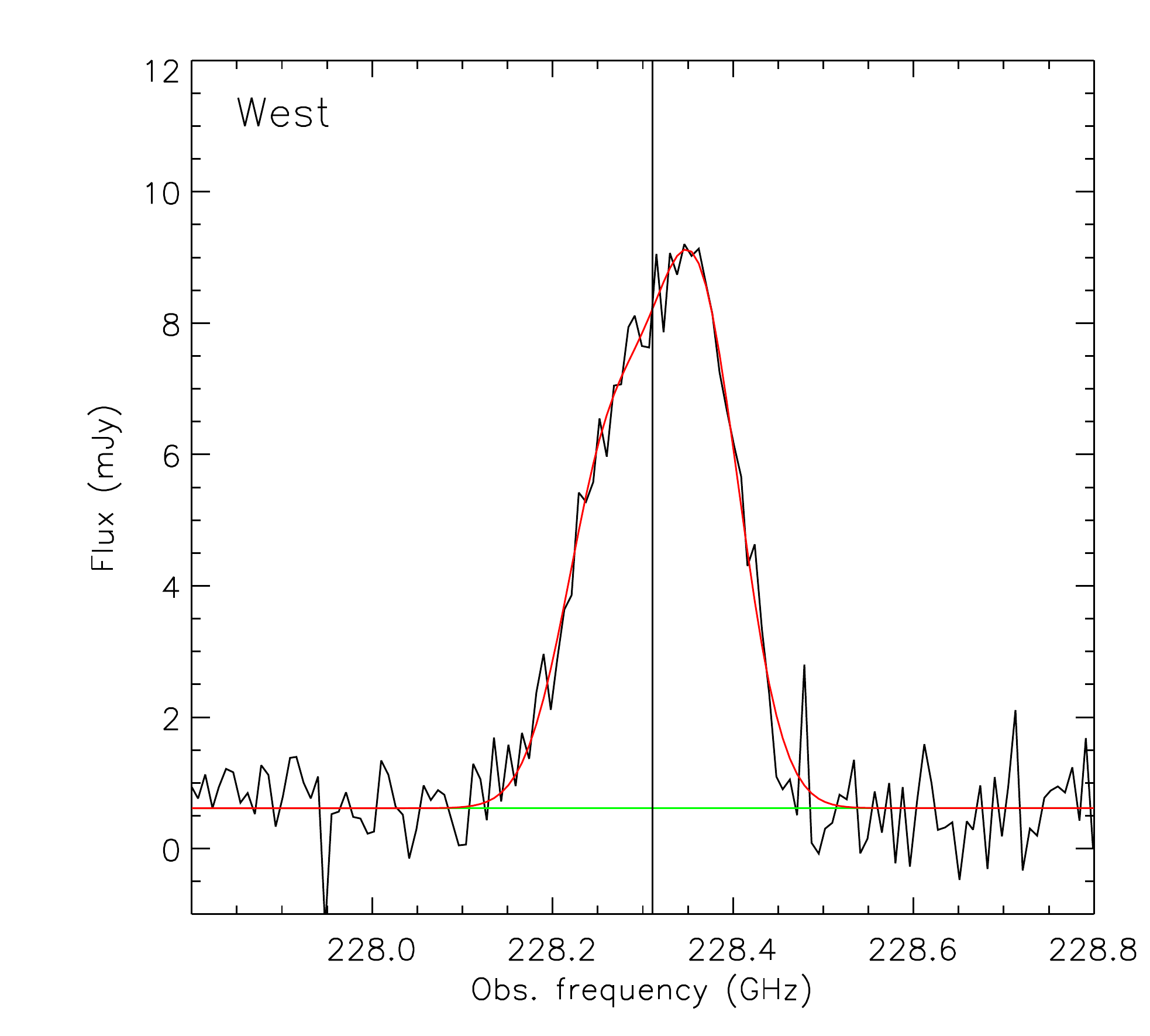}
\caption{CO(5-4) emission-line profile for each galaxy. A best-fit model (in red) is shown for each galaxy based on a double Gaussian function of fixed velocity width with values given in the text and Table~\ref{tab:source_prop}. The vertical line indicates the line centroid used to determine the redshift of each component. The horizontal line (in green) indicates the fit to the continuum that is included in the overall fit to the line profile.}
\label{fig:linefits}
\end{figure}

Using the high-resolution data, CO(5 - 4) emission from PACS-787 is detected at high significance as shown in Figure~\ref{fig:alma}. As opposed to the single source seen at lower spatial resolution (Figure~\ref{fig:lowres}), we now detect two distinct galaxies that each have bright CO(5 - 4) emission. The dual nature of the source at long wavelengths is also evident in the continuum at 1.28 mm (Fig.~\ref{fig:alma}, top-right panel) and in the radio (bottom-right panel) at 3 GHz \citep{Smolcic2017}. The galaxies are clearly undergoing an interaction or merger as evidence by their close proximity both on the sky and along the line of sight with the two redshifts, based on the centroid of the CO emission (Figure~\ref{fig:linefits}), being very close to each other ($z_{East}=1.5256$; $z_{West}=1.5240$), corresponding to a radial velocity difference of $\sim400$ km s$^{-1}$. In further support of a merger scenario, we see signs of low-level extended CO(5 - 4) emission (Section~\ref{sec:diffuse}) and diffuse stellar emission in the HST/WFC3 F140W image (Figure~\ref{fig:images}), both evidence for a likely bridge between the two galaxies.

We measure the CO properties of each galaxy separately by fitting the spatial distribution of the emission using the CASA tool `imfit' with an elliptical Gaussian model. This is carried out on an image of the CO emission, generated with the task `immoments', with a channel (i.e., velocity) width encompassing the full line profile ($\Delta$v) for each galaxy as given in Table~\ref{tab:source_prop}. From the centroid of each component, the projected separation of the two galaxies is 1.02" that corresponds to 8.65 kpc at $z=1.5249$. The half-light radii of the two galaxies are similar and $\sim1$ kpc based on the size of the major axis (East: $r_{1/2}$ = 0.106 $\pm 0.006\arcsec$ = 0.90 $\pm$ 0.05 kpc; West: $r_{1/2}$ = 0.141 $\pm$ 0.005$\arcsec$ = 1.19 $\pm$ 0.04 kpc). With respect to the strength of the CO emission, the eastern galaxy has a slightly higher velocity-integrated flux, I$_{CO}^{east}$=2.82$\pm 0.08$ Jy km s$^{-1}$ ($S/N=35$) as compared to the western galaxy (I$_{CO}^{west}$=2.03$\pm 0.10$ Jy km s$^{-1}$; $S/N=20$). Considering the different velocity intervals over which the line is measured (Sec.~\ref{text:kinematics}), the specific fluxes are S$_{CO}^{east}$=3.77$\pm$0.11 mJy for the eastern and S$_{CO}^{west}$=6.00$\pm$0.30 mJy for the western components. Based on these integrated fluxes, the emission-line luminosities are calculated as follows \citep{Carilli2013}.  

\begin{eqnarray}
L^{\prime}_{\rm CO(5-4)} & = &3.25\times 10^7 S_{\rm CO(5-4)}\Delta \textit{v}~ \nu^{-2}_{\rm obs}~D^2_L (1+z)^{-3} \nonumber \\
 \end{eqnarray}

\noindent We find high CO luminosities ($L^{\prime}_{\rm CO~5-4}$) of $1.3\times10^{10}$ (East) and $9.5\times10^{9}$ K km s$^{-1}$ pc$^2$ (West). The sum of the CO(5 - 4) luminosity for both galaxies is close to the expected value of $L^{\prime}_{\rm CO~5-4}=2.0\times10^{10}$ K km s$^{-1}$ pc$^2$ based on the relation ($L_{CO5-4}$ vs. $L_{IR}$) from \citet{Daddi2015} and the total SFR of 991 M$_{\odot}$ yr$^{-1}$.

\subsection{Gas kinematics}
\label{text:kinematics}

\begin{figure}
\epsscale{1}
\plotone{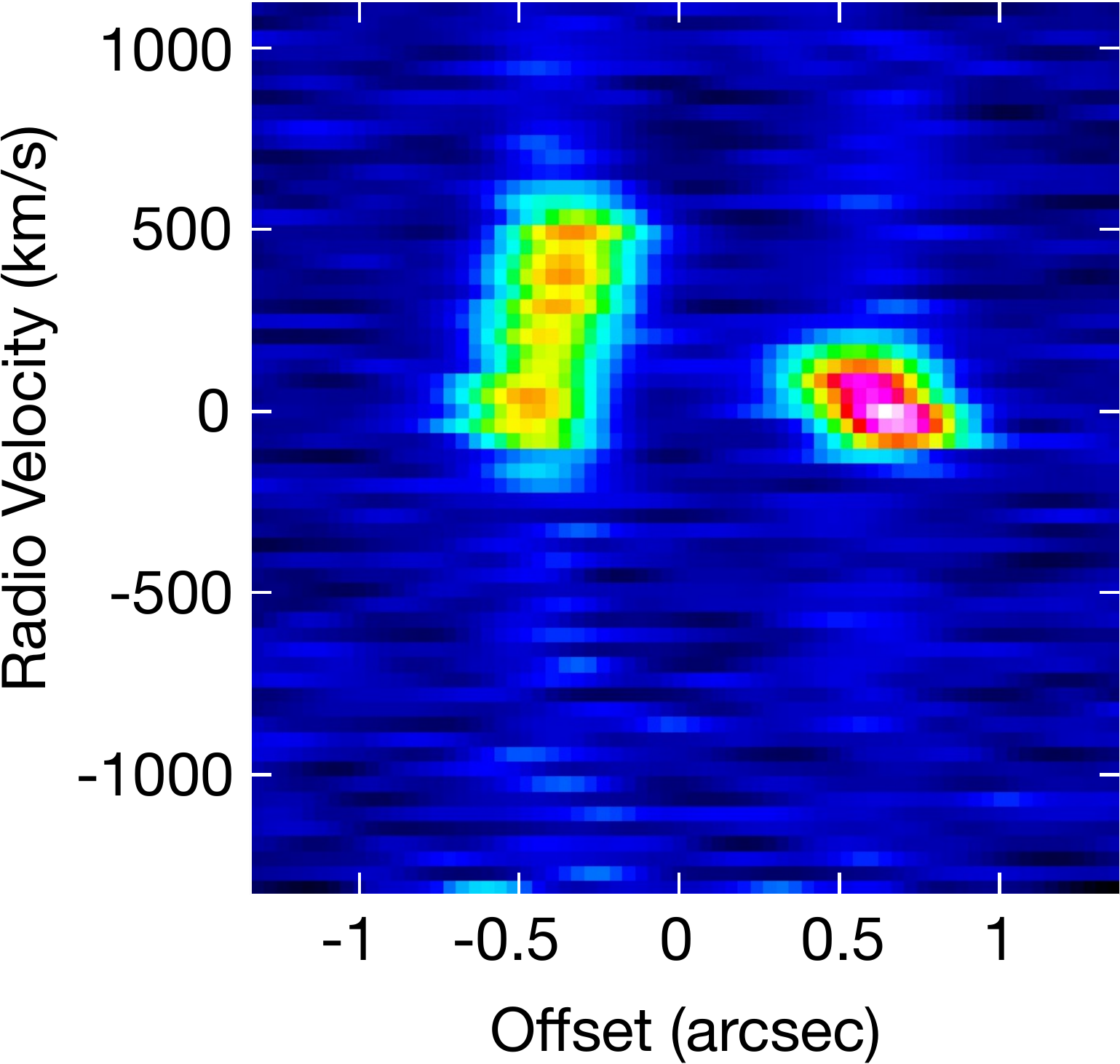}
\caption{Position - velocity map of the CO(5 - 4) emission. The image was constructed using a horizontal spatial region (i.e., collapsing the image along the declination axis) with a width of 5 pixels. The velocity scale is set to zero on the peak of the western source and positive velocities represent red-shifted emission. Faint continuum emission is seen for each galaxy across all velocity channels.}
\label{fig:vel2}
\end{figure}

\begin{figure}
\epsscale{1.2}
\plotone{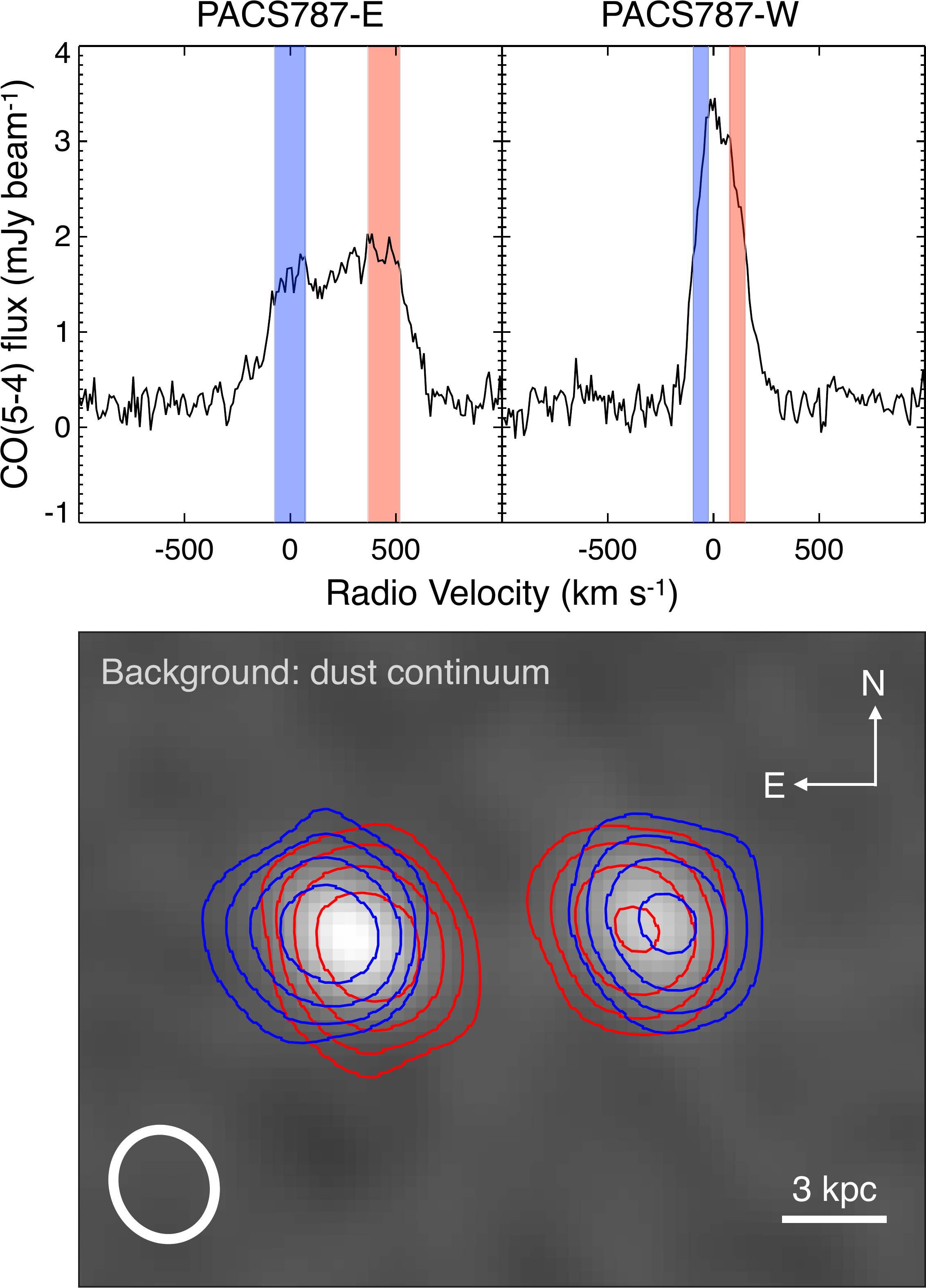}
\caption{The CO velocity profile (top panel) of each galaxy exhibits a double-peaked profile indicative of coherent disk rotation, likely counter-rotating as illustrated by a comparison of the images of the CO emission both red- and blueward from the line center as overlaid on the 1.28 mm continuum image (grey scale; bottom panel). The offsets between the centroids of the red and blue emission for each galaxy are highly significant (East: $\Delta r = 152\pm14$ milliarcsec; West: $\Delta r = 110\pm16$ milliarcsec).}
\label{fig:vel1}
\end{figure}

There is much interest in determining the impact of a major merger on the kinematics of the gas that is initially confined to molecular disks in isolated gas-rich galaxies \citep[e.g.,][]{Hopkins2013}. PACS-787 provides an opportunity to study the gas kinematics in an early stage of such a merger. In fact, we find evidence for the CO emission-line profile for both galaxies to be consistent with a disk in rotation, especially that of the eastern galaxy with an apparent double horn (Figure~\ref{fig:linefits}, top panel). Furthermore, there is a clear relation between position and velocity of the CO emission for both galaxies as shown in Figure~\ref{fig:vel2}. To quantify the significance of the offset between position and velocity, we generate two images for each galaxy in narrow velocity intervals both redward and blueward of the line center as indicated in the top panels of Figure~\ref{fig:vel1}. As expected for resolved emission if undergoing rotation, we find a significant displacement in the spatial location of the blue and red velocity maps for each galaxy (Fig.~\ref{fig:vel1}, bottom panels). For each galaxy, we measure the offsets between red and blue components to be $\Delta r = 152\pm14$ milliarcsec ($1.29\pm0.12$ kpc) and $\Delta r = 110\pm16$ milliarcsec ($0.93\pm0.14$ kpc) for the East and West galaxies, respectively. The difference in the location of the blue and red components for both galaxies indicate a scenario with the disks likely in counter-rotation (as further discussed in Section~\ref{text:discussion}). 

It is also evident that the CO emission-line profiles of the two galaxies are noticeably different in width (e.g., Fig.~\ref{fig:vel1}); the emission line of the western galaxy is significantly narrower than its eastern counterpart. We perform a fit to the emission line to determine the full-width velocity at half maximum ($v_{FWHM}$) of each galaxy with a double Gaussian model of a single fixed velocity width for each component (Figure~\ref{fig:linefits}). The centroid of each Gaussian component is allowed to vary. For both CO emission lines, we find acceptable fits to the line profile in each case (West: $v_{FWHM}= 271.7\pm82.0$ km s$^{-1}$, East:  $v_{FWHM}= 688.4\pm 88.8$ km s$^{-1}$). This leaves little room for additional components such as due to an outflow. As discussed in Section~\ref{text:inclination}, we attribute the different velocity widths to be due to a difference in inclination angle based on the axis ratio of an elliptical Gaussian fit to the CO surface brightness with the western galaxy being closer to face-on than the galaxy to the east.

\subsection{Continuum emission}

For both galaxies, we detect the continuum with a high level of significance (East: $S/N=25$; West: $S/N=17$) at the observed wavelength of 1.276 mm ($\nu=235.0$ GHz). In the top-right panel of Figure~\ref{fig:alma}, we display an image of the two galaxies for which all the spectral channels, devoid of CO line emission, have been summed. The continuum emission is co-spatial with the CO(5 - 4) emission. For the western galaxy, we find similar fit parameters to the surface brightness using an elliptical Gaussian model ($r_{1/2}^{West}=0.14\pm0.02\arcsec$). The eastern continuum source is unresolved. The integrated flux density of the East (West) components are $0.75\pm0.03$ ($0.68\pm0.04$) mJy, respectively. In subsequent sections, we use the continuum emission to estimate the dust mass, dust-to-gas ratio, and SFR for each galaxy.

\section{Molecular gas mass through a dynamical assessment}

We measure the dynamical mass using a method widely employed in the literature \citep[e.g.,][]{Tacconi2008,Daddi2010a,Tan2014, Tadaki2017} to assess the molecular gas content, independent of the CO luminosity or mm continuum. This allows us to evaluate the value of the factor to convert CO luminosity to molecular gas mass (Section~\ref{text:aco}) that is appropriate for our starburst sample and possibly for high-z starbursts in general. This is accomplished by considering the contribution to the dynamical mass (within a half-light radius) from stars, molecular gas, and dark matter. As a novelty, we incorporate a correction (i.e., scale) factor ($f_c$) to account for gas motions, out of equilibrium, that can falsely alter the measured dynamical masses due to turbulence, dispersion, clumpiness, and inflows/outflows (Sec.~\ref{text:simulations}). We measure the dynamical mass, for each galaxy separately, within a half-light radius ($r_{1/2}$) for the case of a thin-disk rotator \citep{Neri2001} as given here:

\begin{equation}
M_{\rm dyn}(r<r_{\rm 1/2}) = f_c\times\frac{r_{1/2}\times(v_{FWHM}/2)^2}{G~sin^2i}
\label{eq:mdyn}
\end{equation}

\noindent where $v_{FWHM}$ is the observed FWHM of the CO(5 - 4) line profile and $r_{1/2}$ is the half-light radius, as reported above and in Table~\ref{tab:source_prop}. We acknowledge that the thin-disk approximation may be a poor choice for high redshift galaxies which are characterized by high velocity dispersion. Therefore, our estimates of $M_{\rm dyn}$ may be somewhat underestimated, and so would be the gas mass (though part of this effect is included in the $f_c$ factor). 

\subsection{Disk inclination angle}
\label{text:inclination}

The disk inclination angle $i$ is estimated from the output parameters of an elliptical Gaussian model fit to the spatial distribution of the CO(5 - 4) emission. We measure the value of $i$ and its uncertainty from the ratio of the size of the minor to major axis. For example, the CO emission for the western galaxy has a major axis (FWHM) of $0.28\pm0.02\arcsec$ and a minor axis of $0.22\pm0.02\arcsec$. We find the inclination angles to be $55_{-6}^{+5}$ (East) and $38_{-10}^{+7}$ (West) degrees for each respective galaxy. We highlight that these constraints on the inclination angle are afforded by the high S/N detections of the ALMA data. As expected, the difference in the inclination angles between the two nuclei concurs with the difference in the velocity widths of the CO lines with the western component closer to being face-on than the eastern nuclei. Remarkably, the differences in inclination of the gas distribution between the two galaxies agrees with the different shapes of the stellar light from each galaxy in the HST WFC3 image (Figure~\ref{fig:images}).

\subsection{Evaluation of $f_c$ using hydrodynamic simulations}
\label{text:simulations}

We use available simulations of gas-rich galaxy mergers \citep{Fensch2017} with similar properties to PACS-787 to determine how well our method to estimate the dynamical mass represents the true dynamical mass since various physical effects, including turbulence, velocity dispersion, clumpiness, and inflows (non-equilibrium effects), may be non-negligible. Our simulation sample includes several mergers of galaxies, with initial conditions (dark matter halo concentration, stellar mass distribution, gas fraction) and interaction orbits representative of star-forming galaxies at redshift z = 1 - 2. These simulations have a maximal resolution of 6 pc, making it possible to model very high-density gas up to $\sim10^{5-6}$ cm$^{-3}$, and to explicitly resolve the main sites of star formation. While these simulations predict that the SFR enhancement in gas-rich mergers at high redshift is, on average, weaker than in gas-poor low-redshift mergers, they do include phases with substantial SFR enhancements and deviations from the MS and Kennicutt-Schmidt relation that characterize isolated star-forming galaxies.

\begin{figure}

\epsscale{0.7}
\plotone{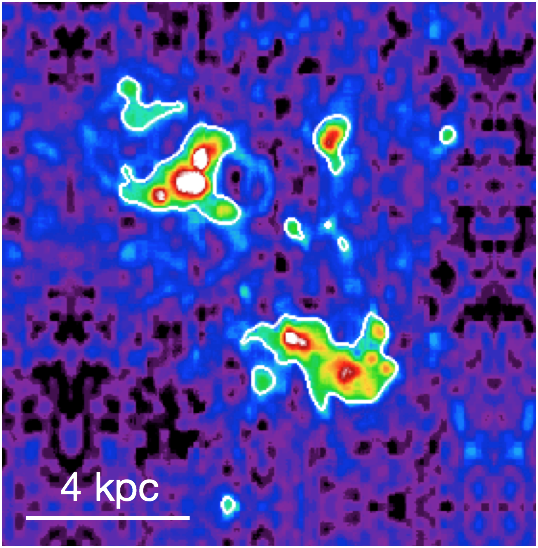}
\epsscale{1}
\plotone{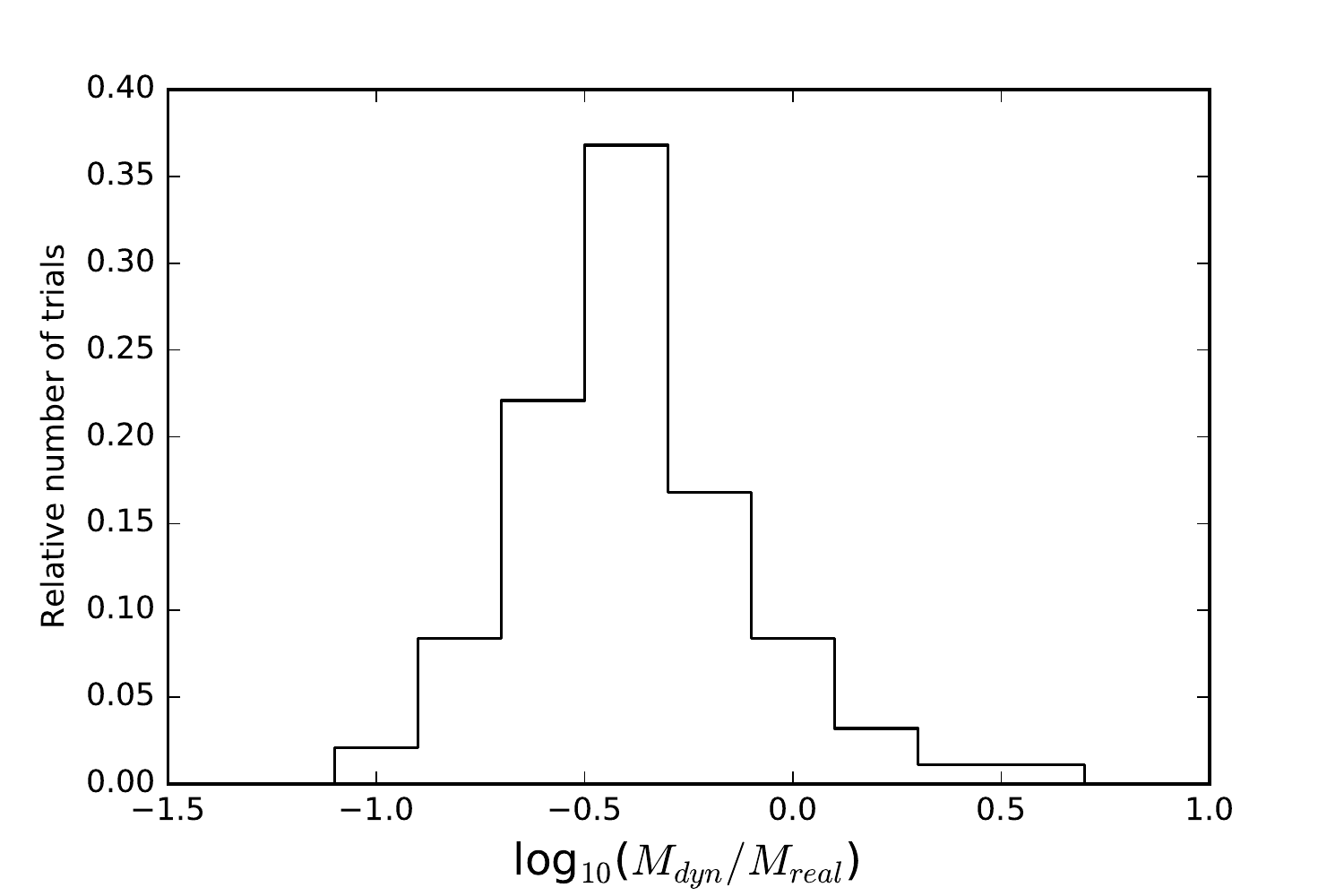}
\caption{Hydrodynamic simulations of high-redshift galaxy mergers \citep{Fensch2017}. $Top$: Example simulation of CO-emitting gas in a galaxy merger with similar properties to PACS-787 at a higher spatial resolution of 0.04" (350 pc at z = 1.5) than our current data. $Bottom$: Ratio of the dynamical mass as measured in Equation~\ref{eq:mdyn} compared to the `true' value from a suite of simulations. The mean of the distribution of trials indicates that $M_{dyn}$ is typically underestimated by a factor of 2.}
\label{fig:hydro}
\end{figure}

We use this sample of simulations to determine the difference between the `observed' and `true' dynamical mass. Specifically, we select 15 snapshots where the merger has triggered a substantial SFR enhancement ($> 3\times$) and generate 10 random projections for each snapshot. We exclude projections where the angular momentum is inclined by less than 20$^{\circ}$ or more than 70$^{\circ}$ from the line-of-sight, because the observed morphology of PACS-787 strongly suggest that the system is not observed as being close to edge-on (the spatial distribution of gas in each galaxy would be highly flattened), and the observed velocity gradients indicate that the system is not observed close to being face-on (the projected velocity amplitudes would be substantially lower). The simulated projections are distributed in an isotropic manner over the allowed range. We post-process the simulation outputs with a dedicated tool \citep{Bournaud2015} that models CO(5 - 4) emission, smoothed to match the spatial resolution of the ALMA observations, and use the mock CO emission line to estimate the dynamical mass as defined in Equation~\ref{eq:mdyn}. We compare to the dynamical mass encompassed in the same region (which is directly known in simulations). We find that our dynamical mass estimates are low by a factor of 2.0 (Figure~\ref{fig:hydro}). The underestimated value results from projection factors (i.e., the projection angle of PACS-787 is unknown and the mock data are analyzed without assuming a known projection angle) and physical effects such as turbulence and non-circular gas motions. We use the average ratio and dispersion (0.2 in log $M_{dyn}/M_{real}$) that encompasses 75\% of the trials shown in Figure~\ref{fig:hydro}, as input ($f_c$) to our dynamical mass estimate.

\subsection{Likelihood distribution of the dynamical mass}

\begin{figure*}
\epsscale{0.8}
\plotone{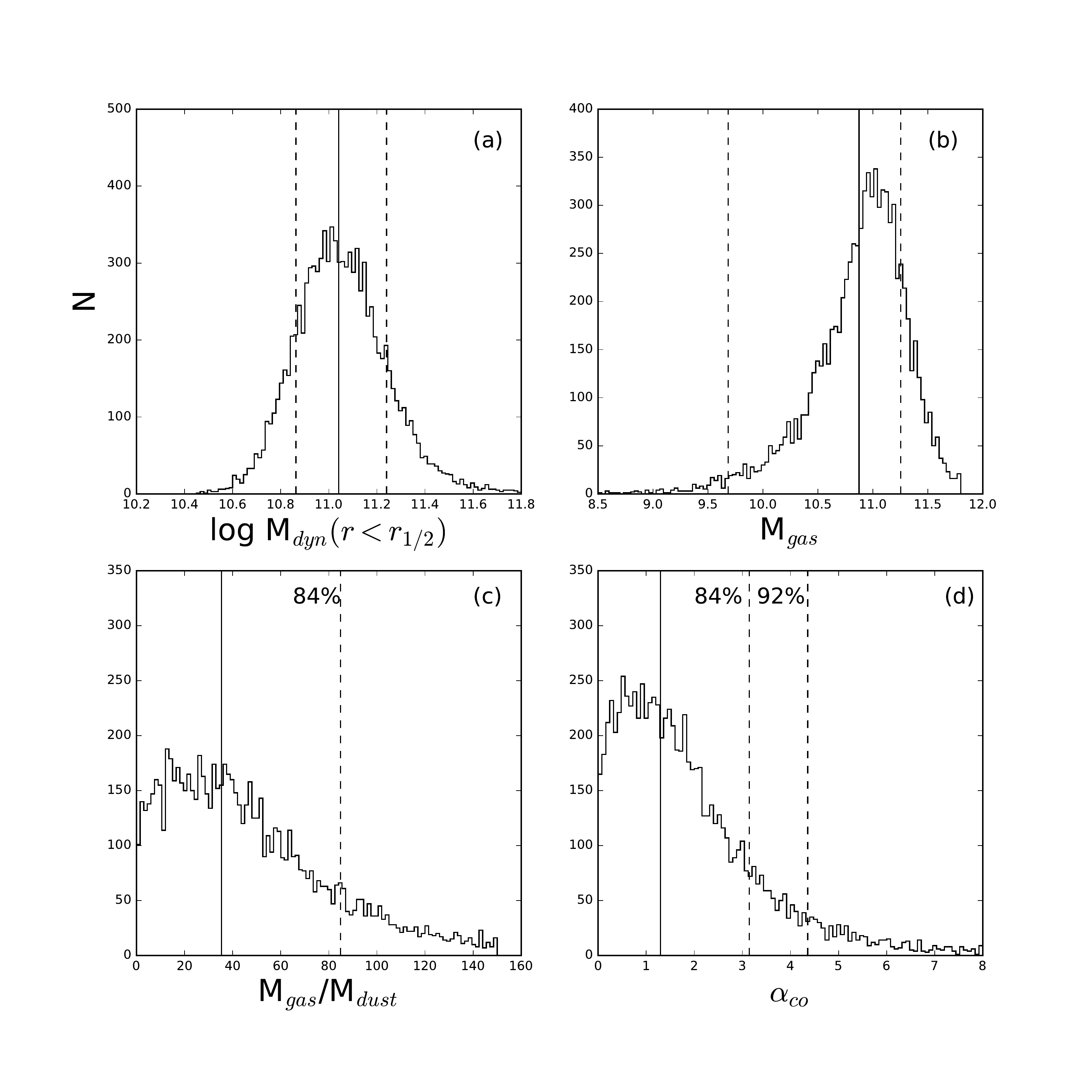}
\caption{Dynamical modeling of the total (i.e., sum of both galaxies) CO emission to derive the (a) dynamical mass within a half-light radius, (b) total gas mass, (c) gas-to-dust ratio, and (d) $\alpha_{CO}$, the ratio of $M_{gas}$ to $L_{CO~1-0}$. Median values and the 68\% regions are indicated by the vertical lines in the top two panels. In the bottom panels, the median value is shown along with vertical dashed lines that indicate the value for which a percentage of trials, as given, are lower.}
\label{fig:mass10k}
\end{figure*}

We calculate the dynamical mass within a half-light radius, as given in Equation~\ref{eq:mdyn}, by generating 10k realizations that incorporate all sources of uncertainty while assuming Gaussian distributions for each parameter with errors (1$\sigma$) provided in Table~\ref{tab:source_prop}. For our purpose of using the dynamical mass to assess the amount of molecular gas mass independent of CO luminosity (see below), we sum the dynamical masses from both galaxies to lessen the effect of random errors on individual measurements. The results of this exercise are shown in Figure~\ref{fig:mass10k}$a$. In addition to the full distribution of trials, we indicate the median dynamical mass ($M_{dyn} (r< r_{1/2} )= 1.1_{-0.4}^{+0.6}\times10^{11}$ M$_{\odot}$) and an interval containing 68\% of the realizations.

\subsection{Contribution of gas to the dynamical mass}

The total gas mass ($M_{gas}$), including both galaxies, is derived from the following equation that relates the different contributions (i.e., $M_{gas}$, $M_{stellar}$, and $M_{dark~matter}$) to the dynamical mass ($M_{dyn}$). 

\begin{equation}
M_{\rm gas}=2\times[M_{\rm dyn}(r<r_{1/2})-M_{dm}(r<r_{1/2})]-M_{stellar}
\label{eq:mgas}
\end{equation}

\noindent 

\noindent {$M_{dyn}$ is determined using Equation~\ref{eq:mdyn} that is applied to our data and simulated observations of galaxy mergers matched to the properties of PACS-787. This ensures that the distribution of $M_{gas}$ includes unphysical results that factor into the assessment of the uncertainties. We assume that the dark matter mass ($M_{dm}$)  is equal to $0.2\times M_{dyn}$ with a dispersion of 0.05, consistent with that of local disk galaxies \citep{DeBlok2008,Genzel2017}. The measurement of the stellar mass is described in Section~\ref{text:mass}. In Figure~\ref{fig:mass10k}$b$, we show 10k realizations to assess the distribution of $M_{gas}$ that has a median gas mass and likelihood distribution (i.e., a 68\% confidence interval) of $M_{gas}=7.5_{-7.1}^{+10.3}\times10^{10}$ M$_{\odot}$. The large estimate of the lower bound of the confidence interval is due to a non-negligible fraction (14\%) of cases that are unphysical due to a dynamical mass that is smaller than the stellar mass. Such solutions are primarily the result of uncertainties on each parameter, in particular, the inclination of the western galaxy (being closer to face-on) or a low correction factor $f_c$. Even so, there are important implications from placing a constraint on the maximum amount of gas that may be present as demonstrated below.

\label{sec:dynamics}

\subsection{$\alpha_{CO}$}
\label{text:aco}

We use the dynamical assessment of $M_{gas}$ for the sum of both galaxies to estimate the median value and upper limits on the ratio of $M_{gas}$ to $L_{CO~1 - 0}^{\prime}$ ($\alpha_{CO}$; units of $M_{\odot}$ (K km s$^{-1}$ pc$^2$)$^{-1}$), a key conversion factor used in the literature to derive gas masses from the CO luminosity \citep[see][for a review]{Bolatto2013}. Currently, there are few constraints on $\alpha_{CO}$ at high redshifts. First, we measure the total CO(1 - 0) luminosity ($L_{CO~1-0}^{\prime}=5.79\pm0.69\times10^{10}$ K km s$^{-1}$ pc$^2$) from our estimate of the CO(2 - 1) luminosity, as reported in \citet{Silverman2018}, and the appropriate conversion factor between CO excitation levels ($L_{CO~2-1}^{\prime} / L_{CO~1-0}^{\prime} = 0.85$; \citealt{Daddi2015}). We confirm that an additional uncertainty ($\sim15\%$) on $L_{CO~2-1}$, due to flux calibration, has no impact on the results due to the larger uncertainty on $M_{gas}$. With 10k realizations (Figure~\ref{fig:mass10k}$d$), we find the median $\alpha_{CO}$ to be 1.3 $M_{\odot}$ (K km s$^{-1}$ pc$^2$)$^{-1}$. Due to the significant number of trials for which our procedure fails to provide a physical result (i.e., a positive gas mass), we cannot place a reliable lower limit on the value of $\alpha_{CO}$. Although, a theoretical lower limit would correspond to $\approx0.3$ $M_{\odot}$ (K km s$^{-1}$ pc$^2$)$^{-1}$ in the optically-thin case \citep[][]{Papadopoulos2012,Bolatto2013}. For the upper limit, we find that 84\% of the trials have a value less than 3.1 $M_{\odot}$ (K km s$^{-1}$ pc$^2$)$^{-1}$ as shown in the figure. Furthermore, 92\% of the realizations have a value less than 4.36 $M_{\odot}$ (K km s$^{-1}$ pc$^2$)$^{-1}$ that is commonly used \citep{Tacconi2018}. If our stellar mass estimates are in reality higher, the amount of gas will be effectively lower, as will $\alpha_{CO}$. Using stellar masses based on a Salpeter as opposed to a Chabrier IMF, we find 96\% of the trials to have $\alpha_{CO} < 4.36$ $M_{\odot}$ (K km s$^{-1}$ pc$^2$)$^{-1}$. Our estimate of $\alpha_{CO}$ given here as 1.3 $M_{\odot}$ (K km s$^{-1}$ pc$^2$)$^{-1}$ is used for subsequent estimates of the total molecular gas mass for each individual galaxy in PACS-787 and the full sample of starbursts having CO(2 - 1) luminosities in \citet{Silverman2018}.

In Figure~\ref{fig:aco}, we compare our estimate of $\alpha_{CO}$ for PACS-787 to other measurements in the literature and derived relations with gas-phase metallicity. We include results based on Local Group galaxies \citep{Leroy2011}, MS galaxies with $z\sim0.5-1.5$ \citep{Magdis2012a}, high-z SMGs (i.e., GN 20, SMM J2135-0102, HERMES J105751.1+573027) and local ULIRGs \citep{Downes1998}. The gas-phase metallicities are based on the calibration of \citet{Pettini2004}. In addition, we show analytic relations from \citet{Genzel2015} and \citet{Tacconi2018}. The metallicity measurement for PACS-787 is based on the [NII]/H$\alpha$ ratio reported in Appendix~\ref{sec:AGN}. From the Figure, it is evident that PACS-787 has a value of $\alpha_{CO}$ closer to that of local ULIRGs and dissimilar to the value of 4.36 $M_{\odot}$ (K km s$^{-1}$ pc$^2$)$^{-1}$ as expected from published relations \citep[e.g., ][]{Tacconi2018}. As discussed in Appendix~\ref{sec:AGN}, there may be an AGN that contributes to the emission line flux, in particular [NII] thus we have placed an arrow indicating that the metallicity may be lower. This has no effect on our conclusions since a lower metallicity would result in a further discrepancy with the higher value of the conversion factor (i.e., 4.36). 

\begin{figure}
\epsscale{1.3}
\plotone{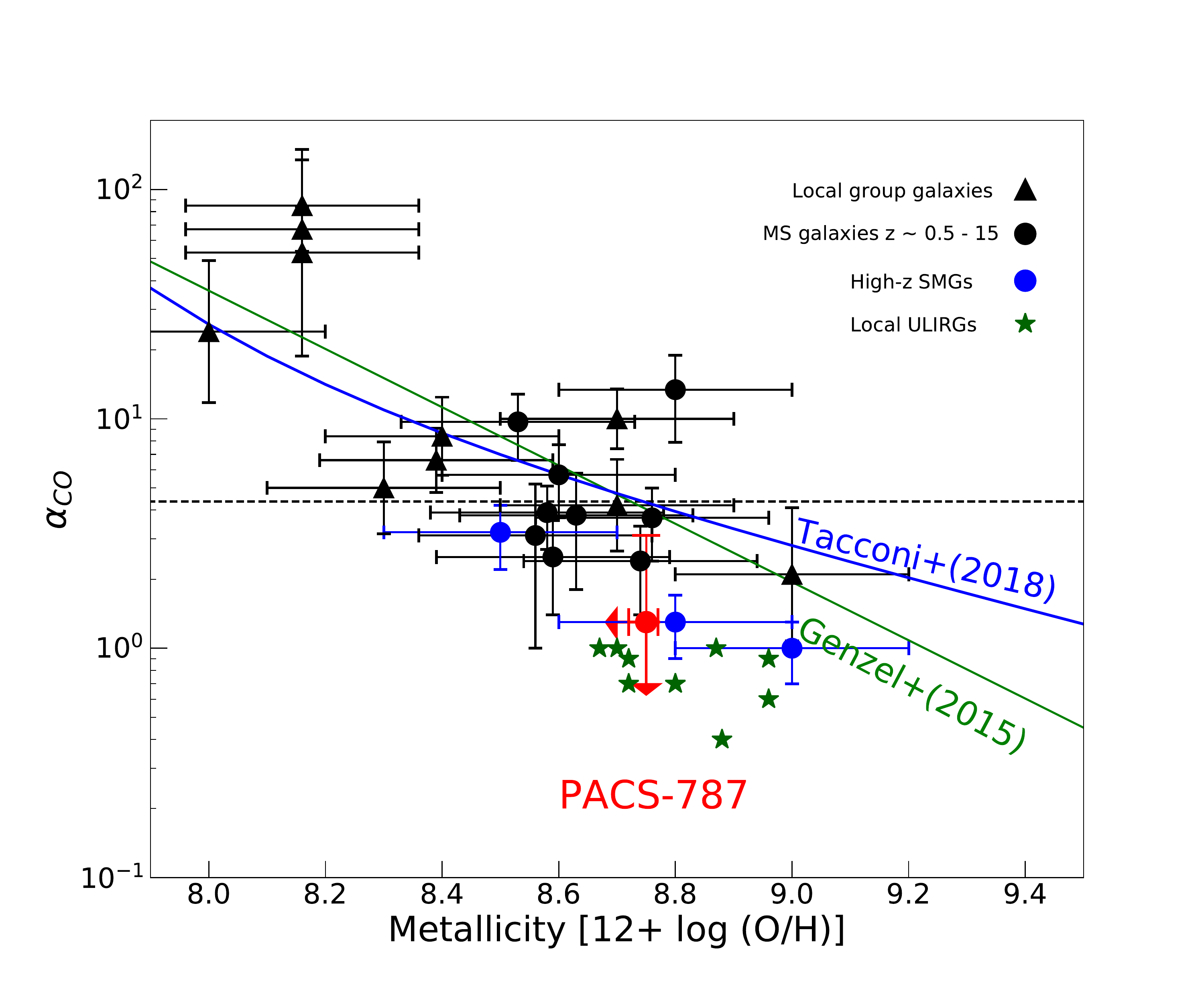}
\caption{$\alpha_{CO}$ as a function of gas-phase metallicity with PACS-787 indicated in red and all other reference samples and relations as described in the text.}
\label{fig:aco}
\end{figure}

\subsection{Dust-to-gas ratio}
\label{text:gd}

Using the total dust mass (Section~\ref{text:sfr}), we estimate the likely value of the gas-to-dust (G/D) ratio to be 35 based on the median of 10k trials (Figure~\ref{fig:mass10k}$c$. On the upper limit, we find that 84\% of the trials have $G/D < 83$. Of course, the gas mass, hence G/D, has considerable uncertainty coming from the difference ($M_{dyn}-M_{stellar} - M_{DM}$) of comparable quantities, each affected by sizable errors. The 10k realizations are meant to incorporate all known uncertainties. In any case, this range of the G/D ratio is consistent that from \citet{Magdis2012a} given its high total stellar mass that likely indicates a high gas-phase metallicity based on the M-Z relation at $z\sim1.5$ from \citet{Kashino2017}. In fact, the [NII]/H$\alpha$ ratio shown in Figure~\ref{fig:agn} could provide evidence for high-metallicity gas. Although, we cannot rule out the influence of a weak AGN on the line ratio. 

\section{Derived physical properties for each galaxy}

\subsection{Gas masses}
\label{co_mass}

Preferably, the gas mass for each individual galaxy should be determined from a dynamical assessment based on their respective CO(5-4) emission. However, there are limitations with respect to this data set, when considering each galaxy separately, due to random effects that we are trying to mitigate as discussed above; thus, for this exercise, we determine the amount of molecular gas in each individual galaxy based on the CO emission which is calibrated by an independent analysis through dynamical modeling (Section~\ref{sec:dynamics}) to assess the conversion factor $\alpha_{CO}$. The total gas mass ($M_{gas}=\alpha_{CO} \times L_{CO~1-0}^{\prime}$) is based on the CO(2 - 1) emission for the two blended galaxies, the conversion factor to CO(1 - 0) luminosity, and the value of $\alpha_{CO}$ to be 1.3 $M_{\odot}$ (K km s$^{-1}$ pc$^2$)$^{-1}$. This is equivalent to using the median total gas mass based on the dynamical argument given above while assessing the error on $M_{gas}$ to reflect the error on CO(2-1) luminosity and not on the uncertainty of $\alpha_{CO}$. The contribution from each galaxy, to the total gas mass is then determined based on the ratio of the CO(5 - 4) emission that is cleanly separated between the two components (Figure~\ref{fig:alma}). This assumes that the excitation states of the two galaxies are the same. As reported in Table~\ref{tab:source_prop}, we find high gas masses for both galaxies: $4.4\pm0.6\times10^{10}$ M$_{\odot}$ (East) and $3.2\pm0.4\times10^{10}$ M$_{\odot}$ (West). With these gas masses, we further estimate the gas fraction $f_{gas}$ to be $70_{-11}^{+8}~(28_{-6}^{+7})~\%$ for the East (West) components respectively as given by Equation~\ref{eq:gasfrac}.

\begin{equation}
\label{eq:gasfrac}
f_{gas}= \frac{M_{gas}}{M_{gas}+M_{stellar}}
\end{equation}

\subsection{Star formation rates and gas consumption timescales}
\label{text:split_sfr}

With each galaxy detected separately by ALMA, HST, VLA and UltraVISTA, we assess the impact of the merger on the SFR for the individual components. We start from the total SFR for both galaxies ($991_{-87}^{+96}$ M$_{\odot}$ yr$^{-1}$) based on $Herschel$ photometry (Section~\ref{text:sfr}). We then determine the SFRs for each galaxy separately (East: $515\pm56$ M$_{\odot}$ yr$^{-1}$; West: $476\pm47$ M$_{\odot}$ yr$^{-1}$) under the assumption that the ratio ($1.10\pm0.08$) of the continuum emission at $\lambda_{rest}=505.5~\mu$m, detected by ALMA, reflects the ratio in SFR. This assumes the dust species and temperature of both galaxies are the same thus have similar SEDs. 

As an independent check, we measure the SFR from the radio flux at 3 GHz \citep{Smolcic2017} that is observed at spatial resolution sufficient to detect each galaxy as shown in the bottom right panel of Figure~\ref{fig:alma}. It is interesting to note that the flux ratio of the 3 GHz emission between the eastern and western galaxy is 1.35, closer to the flux ratio of $L_{\rm CO(5-4)}$ (1.39) than the 1.27mm continuum. These differences are small with respect to the goals of this study so are of not much concern. Even so, we convert the 3 GHz flux to the rest-frame luminosity at 1.4 GHz based on a power-law model with a spectral index $\alpha=-0.7$. Using the radio-SFR relation given in \citet{Bell2003} and applying a conversion to a Chabrier IMF, we find SFRs (East: 438 M$_{\odot}$ yr$^{-1}$; West: 323 M$_{\odot}$ yr$^{-1}$) in fairly good agreement with the estimates given above, including the western galaxy that may have a weak AGN based on a $Chandra$ X-ray detection with very few X-ray counts (Appendix~\ref{sec:AGN}). Therefore, we do not find any evidence for the AGN contributing to the radio luminosity.

\begin{figure}
\epsscale{1.2}
\plotone{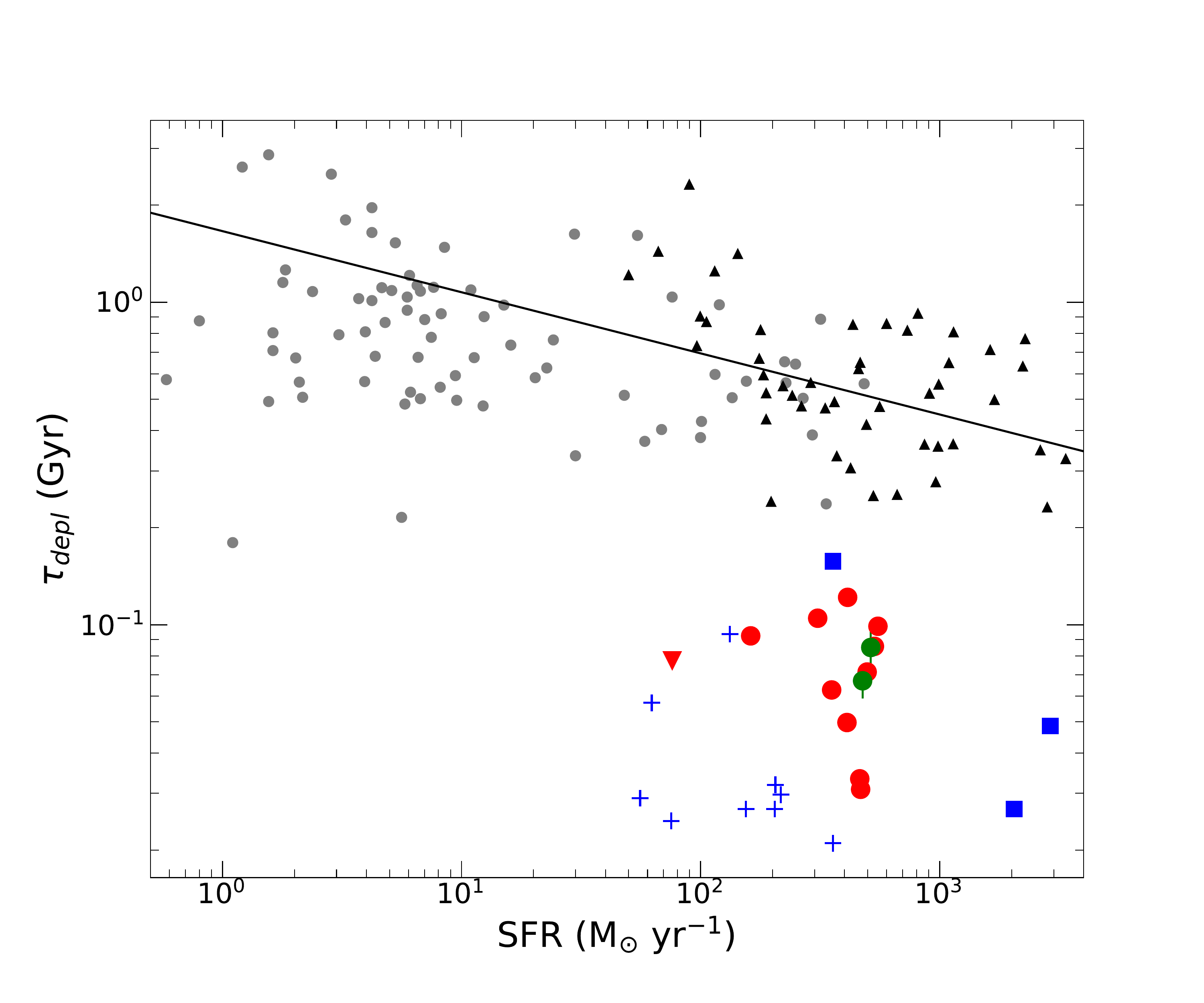}
\caption{Gas depletion time ($\tau_{depl}=M_{gas}/SFR$) of the individual galaxies associated with PACS-787 as a function of their SFR (green symbols) both with 1$\sigma$ errors as indicated. Slanted black line is the best-fit relation from \citet{Sargent2014}. Red (filled) data points mark our other 11 high-z starbursts as described in \citet{Silverman2018}. The inverted triangle represents a 3$\sigma$ upper limit. Small grey circles (black triangles) represent SF MS galaxies at $z<1$ ($>1$) as compiled by \citet{Sargent2014} that include samples from HERACLES \citep{Leroy2013}, COLD GASS \citep{Saintonge2011}, THINGS \citep{Walter2008} and additional higher redshift studies \citep[e.g.,][]{Geach2011,Daddi2010a,Daddi2010b,Magdis2012b,Tacconi2013,Magdis2017}. The blue symbols show the local ULIRGs (crosses; \citealt{Solomon1997}) and additional high-z starbursts (filled blue squares).}
\label{fig:depl_time}
\end{figure}

\begin{figure*}
\epsscale{1}
\plotone{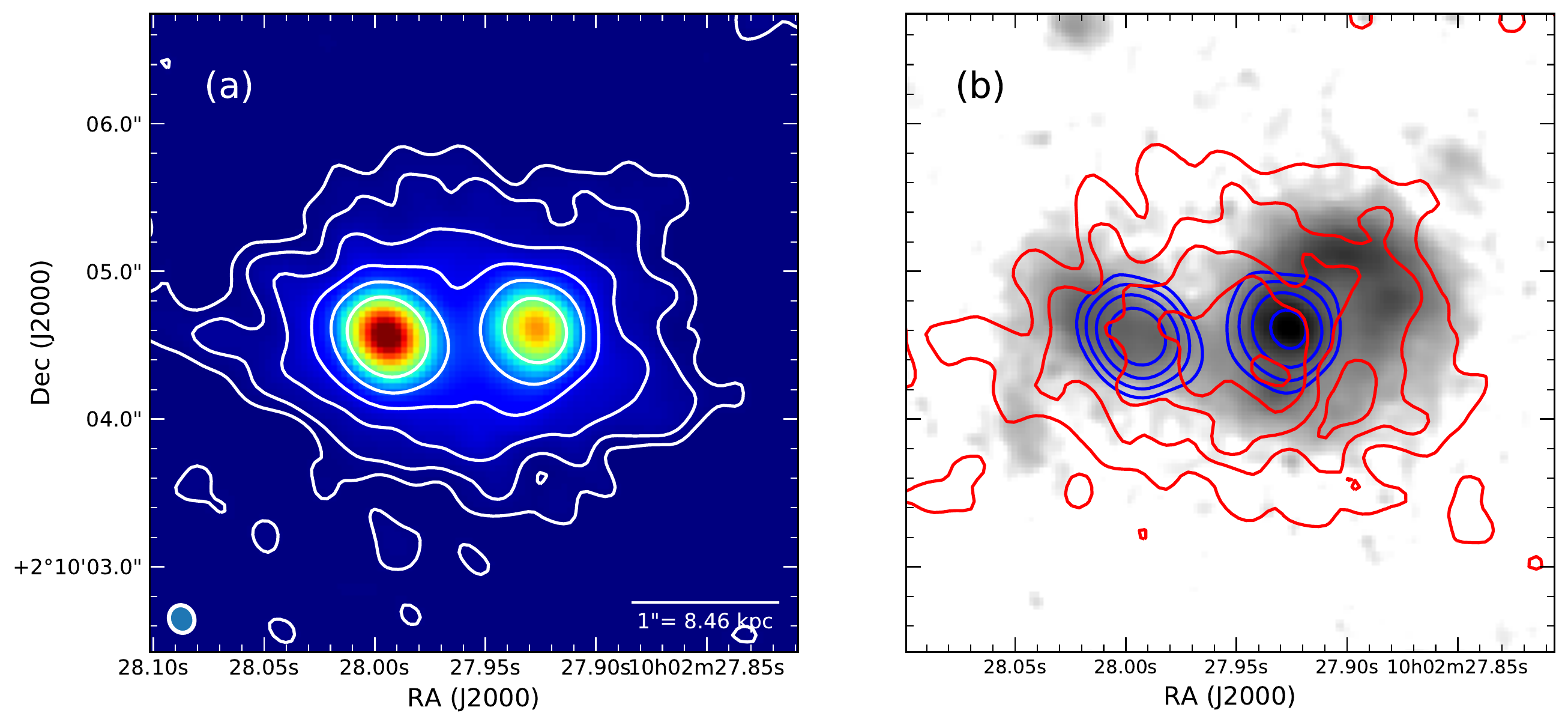}
\epsscale{1.2}
\plotone{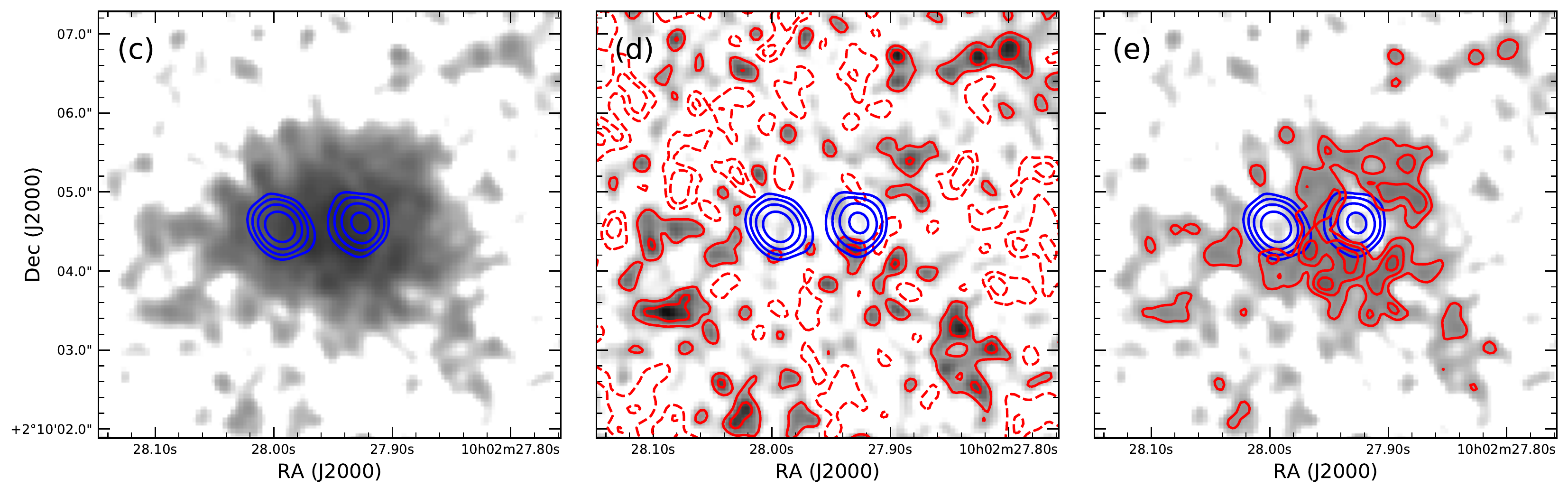}
\caption{Diffuse CO(5-4) emission. (a) ALMA CO(5-4) image generated from the combined low and high resolution data sets over a velocity interval (667.59 km s$^{-1}$; channels 120 - 185) that encompasses the full extent of the CO line for both galaxies. The white contours indicate the CO brightness level at 1, 2, 4, 8, 16, $32\times \sigma_{rms}$ where $\sigma_{rms}=4.67\times10^{-5}$ Jy beam$^{-1}$. (b) After removing the CO emission from both galaxies (based on a three-component fit - see text for details) with their location marked by the blue contours, residual emission (i.e., diffuse gas; red contours at 2, 4, 6, 8, 10$\times\sigma_{rms}$) is present with a bright concentration (i.e., a bridge) between the two nuclei. The diffuse emission supports the scenario that an earlier collision likely occurred at a smaller physical separation. The underlying HST/WFC3 image, shown in grey with a log scaling, displays stellar emission between the two galaxies and co-spatial with the diffuse CO emission. (c) Grey scale image of the diffuse component, fit as a third component, with log scaling, (d) Residual emission after removing all three components (red contours at -2 (dashed), -1, 1 (solid), 2$\times\sigma_{rms}$). (e) Residual emission after subtracting only two components (both galaxies) based on the two-component fit. Contours are shown at 2, 3, $4\times \sigma_{rms}$.}
\label{fig:diffuse}
\end{figure*}

Recently, \citet{Daddi2015} have demonstrated that the integrated CO(5 - 4) luminosity can be used as an indirect tracer of the SFR. As a check on the SFRs measured above, the $L^{\prime}_{\rm CO(5-4)}$ for each galaxy is converted to SFR through its correlation with $L_{TIR}$ in Equation 4 of the aforementioned paper. We find that SFRs are, in general, in very good agreement with those reported above. The western component is essentially the same while the eastern nucleus is sightly higher by 26\%. Each of these methods based on the ALMA fluxes, either the integrated line emission or the continuum, indicate high levels of star formation ($>300$ M$_{\odot}$ yr$^{-1}$) in each galaxy.

Using the SFRs based on total $L_{TIR}$ and the ratio of the continuum luminosity at 1.28 mm, we plot the location of each galaxy in relation to the star-forming MS in Figure~\ref{fig:MS}. We find that both nuclei are undergoing a starburst event having SFRs elevated by $\gtrsim4\times$ above the MS given their stellar mass estimates. In particular, the boost in SFR is notably high (15$\times$) for the less massive galaxy, assuming that both galaxies were on the MS prior to the interaction.

With SFRs and gas masses for each galaxy, we calculate the ratio of these as a star formation efficiency (SFE; Equation~\ref{eq:sfe}) with its inverse usually denoted as a gas depletion time scale ($\tau_{depl}$).
\\
\begin{equation}
\label{eq:sfe}
\tau_{depl}= \frac{M_{gas}}{SFR}; SFE=\frac{1}{\tau_{depl}} 
\end{equation}
\\
\noindent For both galaxies, we find short gas depletion time scales ($< 100$ Myr). For instance, the Eastern (Western) galaxy is rapidly depleted its gas reservoir on a mean time scale of $85\pm14$ ($67\pm10$) Myrs. Both of these timescales are shorter than typical MS galaxies at these SFRs (Figure~\ref{fig:depl_time}), in agreement with that obtained from CO(2 - 1) observations of our larger sample of starburst galaxies \citep{Silverman2018} with an assumption on $\alpha_{CO}$ as measured above, and local ULIRGs \citep{Solomon1997}.

\begin{figure}
\epsscale{1}
\plotone{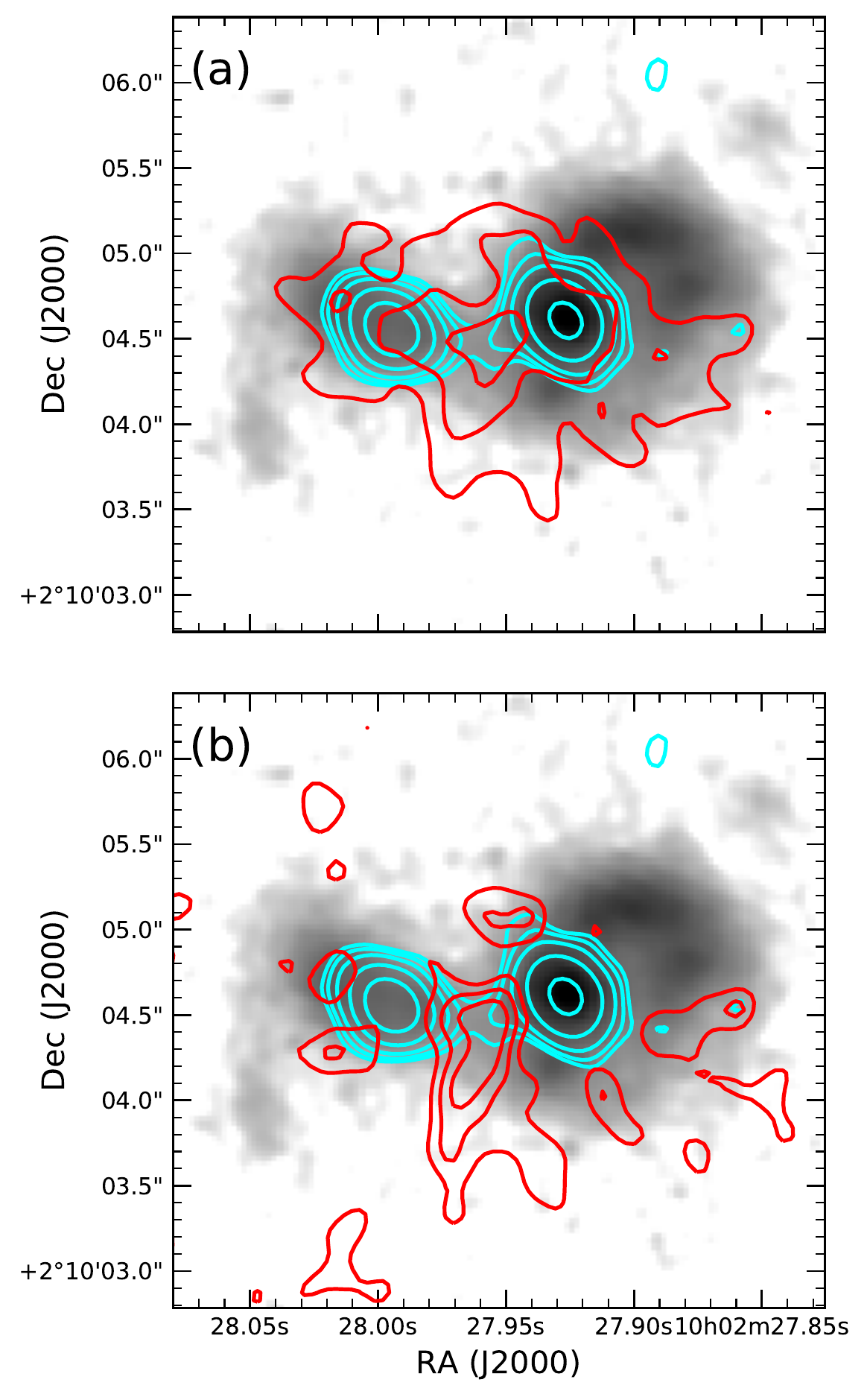}
\caption{Extended continuum emission is present after removing the contribution from both galaxies using either a three- (a) or two-component (b) fit for the individual components. Contours levels in red are 3, 5, 7, 9, 11$\times\sigma_{rms}$ (2, 3, 4, 5$\times\sigma_{rms}$) in panel a (b) where $\sigma_{rms}=1.64\times10^{-5}$ Jy beam$^{-1}$. The HST/WFC3 image is shown in grey. The aqua blue contours indicate the continuum emission from the individual galaxies at levels given in Figure~\ref{fig:alma}.}
\label{fig:diffuse_cont}
\end{figure}

\section{Diffuse CO emission}
\label{sec:diffuse}

We merged the visibilities from the compact and extended ALMA configurations (Band 6), only for this exercise, to gain sensitivity to extended structure and better estimate the total flux for our target galaxies. The combined image ($\Delta$v = 667.6 km s$^{-1}$) is shown in Figure~\ref{fig:diffuse}$a$ with overlying contours that give the significance level of the CO emission. A multi-component fit was performed on the CO(5 - 4) visibility data with an elliptical Gaussian at the position of each of the two galaxies. After removing the two components associated with the galaxies, we found significant residuals with a $5\sigma$ peak, but with substantial extension (hence much higher significance; Fig.~\ref{fig:diffuse}$e$), even though each galaxy may have been over-estimated as shown by the lower residuals (still positive) under each galaxy indicated by the blue contours.

With evidence for an extended component, we re-fit the data allowing for a third Gaussian component and inspected the residuals after subtracting the two galaxies. This resulted in the very extended signal shown by the red contours (Fig.~\ref{fig:diffuse}$b$) and grey scale image (panel c) with minimal ($\lesssim2\sigma$) residuals after subtracting all three components (panel d). With a three-component fit, the significance of the diffuse component is 10$\sigma$ based on the uncertainty of its flux, after marginalizing over the other parameters. As a result of this exercise, we claim that the detection of the extended component of the CO(5-4) emission is robust as further supported by the existence of an extended component with the continuum-only visibilities (Figure~\ref{fig:diffuse_cont}). This extended component appears to contain $\approx20\%$ of the total flux of the system, both in terms of CO(5 - 4) emission and continuum. The diffuse gas/dust component is remarkably co-spatial with the stellar emission as seen in the HST/WFC3 image (Figure~\ref{fig:diffuse}$b$) with the western component being more extended for both the gas and stars. The peak emission is located between both galaxies thus possibly indicative of a bridge connection the two galaxies that may have been tidally stripped during the current or an earlier passage. Furthermore, the strength and morphology of the diffuse component matches very well to the simulation shown in the top panel of Figure~\ref{fig:hydro}.

\section{Discussion and conclusions}
\label{text:discussion}

With our ALMA observations of PACS-787, we can address the physical mechanisms or conditions in the ISM that can produce large enhancements in SFRs since hydrodynamic simulations \citep{DiMatteo2007, Fensch2017} of gas-rich galaxies at high-z generally tend not to exhibit the extreme levels of enhancement as seen in PACS-787. This is because the high gas fractions, in isolated galaxies, already induce high levels of turbulence and the formation of clumps \citep{Bournaud2011} hence high levels of SFRs. PACS-787 is likely an exception that we discovered by selecting, within our sample, one of the most extreme in SFR above the MS for its stellar mass. Furthermore, the two starburst galaxies in PACS-787 are occurring at projected separation of 8.6 kpc. This is unexpected since hydrodynamic simulations usually have the starburst episodes at close to final coalescence. Here we discuss the particular characteristics of PACS-787 for which each may contribute to maximizing the rate of forming stars. 
\\\\
\noindent {\it 1. High gas content:} We find large amounts ($>10^{10}$ M$_{\odot}$) of molecular gas in each galaxy participating in the merger. The eastern galaxy actually has a slightly higher gas mass than stellar mass ($f_{gas}=70_{-11}^{+8}\%$) while the western galaxy has a higher stellar mass and lower gas fraction ($28_{-6}^{+7}\%$). These gas fractions were likely higher prior to the initial interaction since we find $\sim20\%$ of the CO emission on larger scales stripped in the course of the close interaction.  
\\\\
\noindent {\it 2. Near-equal mass merger:} The ratio of the masses (stars + gas) for the two galaxies is 2:1 thus classifying the system as undergoing a major merger. If only considering their stellar mass, the ratio would be reduced to 4:1. The consideration of not only the stellar mass but the molecular gas as well is important in assessing the impact of the merger since hydrodynamical simluations \citep[e.g., ][]{Cox2008} demonstrate that the boost in SFR in galaxy mergers is sensitive to the mass ratio. 
\\\\
\noindent{\it 3. Central gas concentration.} The majority of the CO-emitting gas is confined to regions within a few kpc. The eastern and western galaxies have half-light radius of 0.9 and 1.2 kpc that are similar to local ULIRGs \citep{Solomon1997} and continuum observations with ALMA of high-z submm galaxies \citep{Simpson2015}.
\\\\
\noindent{\it 4. High star formation efficiency.} Both galaxies are highly efficient at forming stars similar to that seen in local ULIRGs. We find very similar gas depletion times of $\sim$70-80 Myrs for the eastern and western galaxies respectively. These time scales indicate a consumption of their molecular gas much shorter than typical MS galaxies at an equivalent SFR (Figure~\ref{fig:depl_time}).  
\\\\
\noindent {\it 5. Orbital configuration:} Even with limited information on the orbital dynamics of the system, we estimate the amount of angular momentum, expressed as $L=m_{gas}\times r_{1/2}\times v$, that may be canceled in the merger using our measurements  in hand for each galaxy: size ($r_{1/2}$), inclination ($i$), velocity ($v=  v_{FWHM} / (2\times sin~i)$), and gas mass ($m_{gas}$). In all four possible orientations for the two disks (counter-rotating vs co-rotating; inclination being positive or negative with respect to the plane of the sky), the component of the total angular momentum in the plane of the sky for each individual galaxy will be opposite to each other. We note that in the co-rotating case, the disks are essentially perpendicular to each other hence there is no distinction from co- or counter-rotation. We find that the difference in angular momentum for the two components (in the plane of the sky) will be $\sim$37\% of the sum of the angular momentum of both galaxies. This is a considerable amount of angular momentum that may be lost and possibly relevant in explaining the high star formation rates. For this exercise, we neglect to consider the total angular momentum of the system.

We highlight that these results are derived from a procedure to assess the amount of molecular gas independent of the CO luminosity. The high signal-to-noise detection of CO(5-4) for each galaxy and the survival of the their kinematic structure as a disk at this stage of the merger allow us to carry out such an analysis. Whether this situation is fortuitous or prevalent throughout the high-z starburst population is yet to be determined. Additional examples are obviously needed given their impact on the use of CO as a tracer of the molecular gas mass at high redshift. We refer the reader to the discussion in \citet{Silverman2018} on the impact of uncertainties on $\alpha_{CO}$ in measuring the molecular gas content for high-z starburst galaxies.

To conclude, the observations presented here of a quintessential starburst induced by the pre-merging interaction of two galaxies at high redshift demonstrate the remarkable capability of ALMA to study such systems in detail. In particular, the molecular gas and dust properties are measured for both galaxies at a stage in the merger prior to final coalescence. As indicative of the interaction, we detect diffuse CO emission, likely tidally-stripped, for the first time at high redshift ($z\gtrsim1$). With sufficient spatial and spectral resolution, we were able to model the kinematics of the system to derive a dynamical mass and account of the mass contribution from the stellar and gas mass with an assumption on the amount of dark matter present. For the first time, we use simulations of galaxy mergers to calibrate the estimate of the dynamical mass which indicate that other attempts may be off by a factor $\sim2$. As described above, this led to constraints on two key parameters ($\alpha_{CO}$ and the G/D ratio) that are poorly established at high redshift. Finally, we expect that future ALMA observations of PACS-787 at a higher spatial resolution will unveil the structure of the gas and its velocity structure on scales ($\sim100$ pc) of giant molecular cloud complexes

\begin{deluxetable*}{lllll}
\tabletypesize{\small}
\tablecaption{PACS-787: Properties of individual components \label{tab:source_prop}}
\tablehead{\colhead{Quantity}&\colhead{East} & \colhead{West}& Units&\colhead{Method}}
\startdata
RA&10:02:27.995 ($\delta=0.004\arcsec$)&10:02:27.927($\delta=0.006\arcsec$)&&Centroid of the continuum emission\\
DEC&+02:10:04.56&+02:10:04.61&\\
z$_{CO}$&1.5256&1.5240&&Redshift; $\delta_z=0.0001$\\
M$_{stellar}$&$1.9_{-0.4}^{+0.8}\times10^{10}$&$8.1_{-1.7}^{+1.9}\times10^{10}$&M$_{\odot}$&Chabrier IMF\\
I$_{CO(5 - 4)}$&2.82$\pm0.08$&2.03$\pm$0.10&Jy km s$^{-1}$\\
$\Delta$v$$ &748.7&338.4&km s$^{-1}$&Velocity range over which I$_{CO(5 - 4)}$ is measured.\\
L$^{\prime}_{\rm CO(5-4)}$ &$1.3\times10^{10}$&$9.5\times10^{9}$&K km s$^{-1}$ pc$^2$\\
$v_{FWHM}$&$688.4\pm88.8$&$271.7\pm82.0$&km s$^{-1}$&CO velocity width at half maximum from Gaussian fit\\
$r_{1/2}$&$0.90\pm0.05$&$1.19\pm0.04$&kpc&CO; half-light radius\\
$i$&$55_{-6}^{+5}$&$38_{-10}^{+7}$&degrees&Disk inclination (CO)\\
f$_{1.3~{\rm mm}}$ &0.75$\pm$0.03&0.68$\pm0.04$&mJy\\
f$_{3GHz}$&70.4$\pm$4.0&52.0$\pm$3.9&$\mu$Jy\\
SFR &$515\pm56$&$476\pm47$&M$_{\odot}$ yr$^{-1}$&\\
$M_{gas}$ &$4.4\pm0.6\times10^{10}$&$3.2\pm0.4\times10^{10}$&M$_{\odot}$\\
$f_{gas}$&$70_{-11}^{+8}$&$28_{-6}^{+7}$&\%\\
$\tau_{depl}$&$85\pm14$&$67\pm10$&Myr\\
\enddata
\end{deluxetable*}

\acknowledgments

\appendix

\section{AGN}
\label{sec:AGN}

PACS-787 may be the host to an Active Galactic Nucleus (AGN), powered by a supermassive black hole. There is very weak hard (2-7 keV) emission detected by $Chandra$ (Figure~\ref{fig:agn}; left panel). This source was not found in previous X-ray analysis \citep{Civano2016} of the COSMOS field, possibly hampered by having only 4 X-ray counts in the 2-7 keV band and at a distance of 11$\arcsec$ to a bright X-ray source to the southeast that has a photometric redshift of 1.27 \citep{Marchesi2016}. Attributing all the X-ray counts within an aperture of 8$\arcsec$ to PACS-787, the AGN would have $L_X\sim2\times 10^{43}$ ergs s$^{-1}$ that may be responsible for the high ionization state of the interstellar medium as indicated by the ratio [NII]/H$\alpha=0.54\pm0.04$ (right panel). Although, we cannot rule out the possibility that PACS-787 has a high metallicity as opposed to an underlying AGN as the source of photoionization. Whether the X-ray detection is real or not, there is no impact on the analysis used in this study that would alter our conclusions.

\begin{figure}
\epsscale{0.4}
\plotone{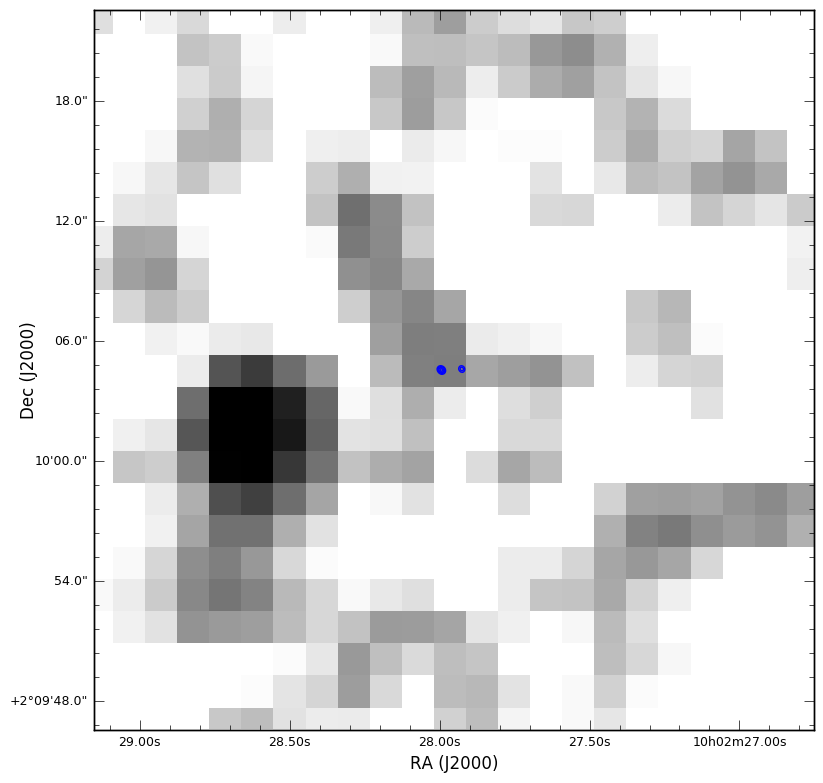}
\epsscale{0.55}
\plotone{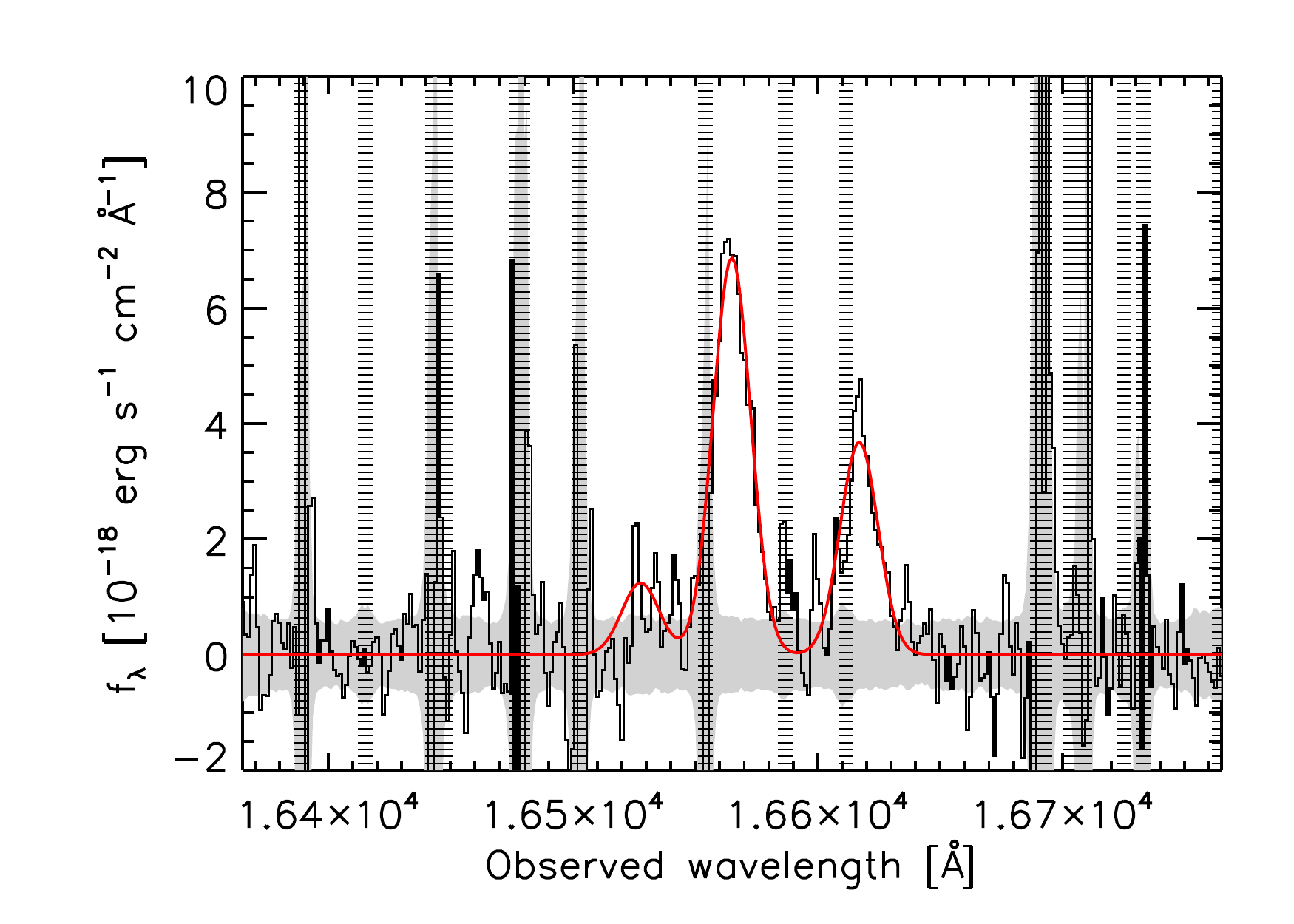}
\caption{Marginal evidence for an accreting supermassive black hole in PACS-787. (Left) Gaussian smoothed (2 pixels) $Chandra$ X-ray image (2-7 keV). The blue spots (at the center of the image) indicate the location of the two components of PACS-787. A faint X-ray source may be close to the position of PACS-787. (Right) Subaru/FMOS near-infrared spectrum of the western galaxy (with the aperture location shown in the bottom right panel of Fig.~\ref{fig:images}). Grey hatched regions mark those wavelengths impacted by OH emission. The red curve is a best-fit Gaussian model to the H$\alpha$ and [NII] emission lines.}
\label{fig:agn}
\end{figure}

\acknowledgments

We are grateful for the support from the regional ALMA ARCs. JDS was supported by the ALMA Japan Research Grant of NAOJ Chile Observatory, NAOJ-ALMA-0127, JSPS KAKENHI Grant Number *JP18H01521*, and the World Premier International Research Center Initiative (WPI Initiative), MEXT, Japan. CM and AR acknowledge support from an INAF PRIN 2012 grant.  W.R. is supported by Thailand Research Fund/Office of the Higher Education Commission Grant Number MRG6080294 and Chulalongkorn University's CUniverse. GEM acknowledges support from the Carlsberg Foundation, the ERC Consolidator Grant funding scheme (project ConTExt, grant num- ber No. 648179), and a research grant (13160) from Villum Fonden. N.A. is supported by the Brain Pool Program, which is funded by the Ministry of Science and ICT through the National Research Foundation of Korea (2018H1D3A2000902). This paper makes use of the following ALMA data: ADS/JAO.ALMA\#2012.1.00952.S, ADS/JAO.ALMA\#2015.1.00861.S and ADS/JAO.ALMA\#2016.1.01426.S. ALMA is a partnership of ESO (representing its member states), NSF (USA) and NINS (Japan), together with NRC (Canada), NSC and ASIAA (Taiwan), and KASI (Republic of Korea), in cooperation with the Republic of Chile. The Joint ALMA Observatory is operated by ESO, AUI/NRAO and NAOJ.




\end{document}